\documentclass[nonacm,sigplan]{acmart}

\usepackage{lipsum}

\usepackage{pdfpages}
\usepackage{bm}
\usepackage{algorithm}
\usepackage{listings}
\usepackage{subfig}
\usepackage{array}
\usepackage{enumitem}
\usepackage{xcolor}
\usepackage{caption}
\usepackage{tikz}
\usepackage{tcolorbox}
\usepackage{graphicx}
\usepackage{textcomp}
\usepackage{algpseudocode}
\usepackage{multirow}
\usepackage{booktabs}
\usepackage{verbatim}
\usepackage{atbegshi}
\usepackage{hyperref}
\usepackage[hyphenbreaks]{breakurl}
\usepackage{titlesec}
\usepackage{pifont}
\usepackage{makecell}
\usepackage{afterpage}
\newlength{\savedtextfloatsep}
\newcommand{\tighttextfloatsepOnce}[1]{%
  \global\savedtextfloatsep=\textfloatsep
  \global\textfloatsep=#1\relax
  \afterpage{\global\textfloatsep=\savedtextfloatsep}%
}
\tcbuselibrary{skins}
\tcbset{
    colback=white,
    colframe=white,
    arc=1mm,
    left=0mm,
    right=0mm,
    top=0mm,
    bottom=0mm,
    parbox=false,
    boxrule=1.2pt,
    borderline={1.2pt}{0pt}{black,dashed}
}
\usepackage{hyperref}
\usepackage[hyphenbreaks]{breakurl}

\definecolor[named]{ACMPurple}{cmyk}{0.55,1,0,0.15}

\newcommand{\colorcite}[1]{\textcolor{ACMPurple}{\cite{#1}}}
\newcommand{\colorref}[1]{\textcolor{ACMPurple}{\ref{#1}}}

\begin{document}

\graphicspath{{figures/}}

\title{DALI: A Workload-Aware Offloading Framework for Efficient MoE Inference on Local PCs}

\author{Zeyu Zhu}
\affiliation{%
  \institution{Institute of Automation, CAS}
  \institution{School of Future Technology, University of Chinese Academy of Sciences}
  \country{Beijing, China}
}
\email{zhuzeyu2021@ia.ac.cn}

\author{Gang Li}
\affiliation{%
  \institution{Institute of Automation, CAS}
  \country{Beijing, China}
}
\email{gang.li@ia.ac.cn}

\author{Peisong Wang}
\affiliation{%
  \institution{Institute of Automation, CAS}
  \country{Beijing, China}
}
\email{peisong.wang@nlpr.ia.ac.cn}

\author{Zitao Mo}
\affiliation{%
  \institution{Institute of Automation, CAS}
  \country{Beijing, China}
}
\email{mozitao2017@ia.ac.cn}

\author{Minnan Pei}
\affiliation{%
  \institution{Institute of Automation, CAS}
  \country{Beijing, China}
}
\email{peiminnan19@mails.ucas.ac.cn}

\author{Zhuoran Song}
\affiliation{%
  \institution{Shanghai Jiao Tong University}
  \country{Shanghai, China}
}
\email{songzhuoran@sjtu.edu.cn}

\author{Xiaoyao Liang}
\affiliation{%
  \institution{Shanghai Jiao Tong University}
  \country{Shanghai, China}
}
\email{liang-xy@cs.sjtu.edu.cn}

\author{Jian Cheng}\authornote{Corresponding author.}
\affiliation{%
  \institution{Institute of Automation, CAS}
  \institution{AiRiA}
  \institution{Maicro.ai}
  \country{Beijing, China}
}
\email{jcheng@nlpr.ia.ac.cn}

\begingroup
\spaceskip=0.98\fontdimen2\font plus 0.95\fontdimen3\font minus 1.15\fontdimen4{abstract}

  Mixture of Experts (MoE) architectures significantly enhance the 
  capacity of LLMs without proportional increases in computation, but at the 
  cost of a vast parameter size. Offloading MoE expert parameters to host 
  memory and leveraging both CPU and GPU computation has recently emerged 
  as a promising direction to support such models on resource-constrained 
  local PC platforms.
  While promising, we notice that existing approaches mismatch the dynamic nature of expert workloads, which 
  leads to three fundamental inefficiencies:
  (1) Static expert assignment causes 
    severe CPU-GPU load imbalance, underutilizing CPU and GPU resources; 
    (2) Existing prefetching techniques fail to 
    accurately predict high-workload experts, leading to costly inaccurate prefetches; 
    (3) GPU cache policies neglect workload dynamics, resulting in poor hit rates   
    and limited effectiveness. To address these challenges, we propose \textbf{DALI}, 
    a workloa\underline{\textbf{D}}-\underline{\textbf{A}}ware 
    off\underline{\textbf{L}}oad\underline{\textbf{I}}ng framework for efficient MoE inference on local PCs.
    To fully utilize hardware resources,
    DALI first dynamically 
    assigns experts to CPU or GPU by modeling assignment as a 0-1 integer 
    optimization problem and solving it efficiently using a \textbf{Greedy Assignment} strategy at runtime. 
    To improve prefetching accuracy, we develop a \textbf{Residual-Based Prefetching} method 
    leveraging inter-layer residual information to accurately predict high-workload 
    experts. Additionally, we introduce a \textbf{Workload-Aware Cache Replacement} policy that 
    exploits temporal correlation in expert activations to improve GPU cache efficiency.
    By evaluating across various MoE models and settings,
    DALI achieves significant speedups in the both prefill and decoding phases over the 
    state-of-the-art offloading frameworks.

\end{abstract}
\endgroup
\maketitle 
\pagestyle{plain} 

\begingroup
\spaceskip=0.95\fontdimen2\font plus 0.95\fontdimen3\font minus 1.15\fontdimen4\font
\section{Introduction}

\captionsetup{font=small}

Recently, \textbf{Mixture of Experts (MoE)} architectures have been widely adopted in 
Large Language Models (LLMs), including {Switch Transformers}~\colorcite{switchtrans}, 
{Mixtral}~\colorcite{jiang2024mixtral}, {DeepSeek}~\colorcite{liu2024deepseek}, and 
{Qwen}~\colorcite{chu2024qwen2}, for their ability to significantly enhance model capacity 
without proportionally increasing computation~\colorcite{rajbhandari2022deepspeed, 
shazeer2017outrageously}. In MoE, a gating function selects a small subset of experts for 
each token, allowing computation to be focused only on the activated experts. While this 
approach improves efficiency, it also substantially increases the total parameter count
~\colorcite{abdin2024phi,dai2024deepseekmoe}, posing significant deployment challenges on 
resource-constrained platforms such as personal computers with memory-limited GPUs.

\begin{table}[t]
  \centering
  \caption{Comparison of Local PC and High-End Server Hardware}
  \label{tab:hw_compare}
  \small
  \setlength{\tabcolsep}{3.5pt} 
  \vspace{-0.2cm}
  \begin{tabular}{ccc}
  \toprule
  \textbf{Feature} & \textbf{Local PC} & \textbf{H100 Server} \\
  \midrule
  GPU Model & RTX 3090/4090 & H100 80GB \\
  GPU Memory & 24--32 GB & 80 GB \\
  GPU Bandwidth & $\sim$936--1000 GB/s & $\sim$3400 GB/s \\
  PCIe Bandwidth 
  & \makecell[c]{PCIe 4.0 $\times$16 \\ \textbf{32 GB/s}} 
  & \makecell[c]{PCIe Gen5 / NVLink \\ \textbf{128--900 GB/s}} \\
  System RAM & 32--128 GB & 256--1024+ GB \\
  System Cost & \$2k--\$5k & \$200k--\$400k+ \\
  \bottomrule
  \end{tabular}
  \vspace{-0.6cm}
\end{table}

\begin{figure*}[t]
  \centering
  \vspace{-0.7cm}
  \includegraphics[scale=0.63,trim=0cm 0cm 0cm 0.0cm,clip]{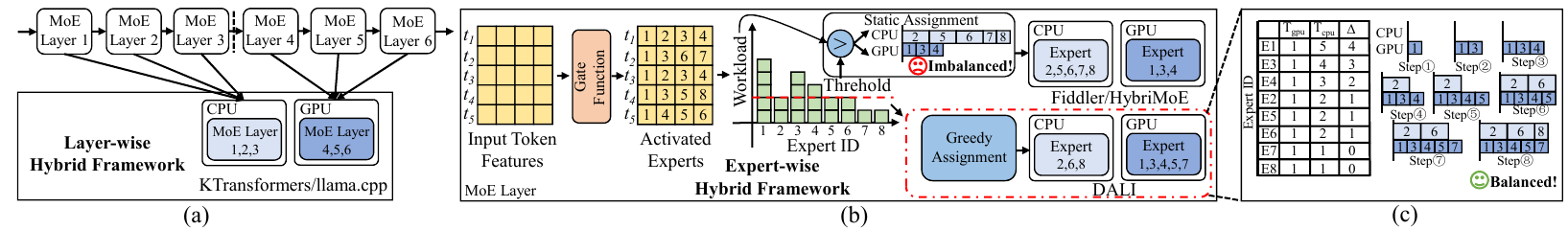}
  \vspace{-0.4cm}
  \caption{Different MoE offloading frameworks. $t_i$ denotes the $i$-th token. 
  DALI dynamically schedules the experts based on the workloads, 
  thus achieving better CPU and GPU balance.}
  \vspace{-0.4cm}
  \label{intro_to_workload}
\end{figure*}

Table~\colorref{tab:hw_compare} highlights the gap between local PCs (e.g., RTX 3090/4090 setups) 
and high-end servers (e.g., NVIDIA H100). H100-based systems offer substantially more 
compute and memory resources, but are prohibitively expensive---a complete system can 
cost over \$200{,}000 and is inaccessible to most users. In contrast, local PCs are far 
more affordable and widely available, yet constrained by limited GPU memory and PCIe 
bandwidth. Therefore, enabling efficient MoE inference on local PCs is a 
critical research problem that can democratize LLM deployment and reduce reliance on 
costly cloud infrastructure.

Offloading is a promising strategy to 
alleviate the memory demands of MoE models by storing expert parameters in 
secondary memory (e.g., DRAM, SSD, HDD), thereby reducing GPU memory usage without 
compromising model expressiveness. Conventional offloading frameworks 
transfer expert weights from CPU to GPU via PCIe after expert activation is 
determined \colorcite{tang2024hobbit, xue2024moe, eliseev2023fast, yi2023edgemoe, 
yu2025fmoe, song2024promoe, he2024expertflow, fang2025accurate, zhang2025daop, 
zhong2024adapmoe, hwang2024pre, cao2025moe}. However, due to limited PCIe bandwidth of local PCs and the 
large size of MoE parameters, they incur considerable inference latency 
and restricts deployment in real-world scenarios.
To reduce PCIe communication overhead, 
recent frameworks \colorcite{llama.cpp, ktransformers, kamahori2024fiddler, zhong2025hybrimoe} 
adopt hybrid CPU-GPU execution and offload expert computation to CPUs, thus 
mitigating
data transfer costs and improving throughput. 
Nevertheless, expert activation is inherently input-dependent in MoE layers, causing 
the token count routed to each expert (i.e., the expert workload) to vary widely across 
inputs. This workload dynamism introduces three critical challenges for existing hybrid 
offloading frameworks:

\textbf{Underutilization of heterogeneous computational resources.} 
As shown in Figure~\colorref{intro_to_workload}\textcolor{ACMPurple}{a}, 
llama.cpp \colorcite{llama.cpp} and Ktransformers \colorcite{chen2025ktransformers} 
assign MoE layers to either the CPU or GPU, 
executing each layer on the device where the parameters reside
(referred to as \textit{\textbf{layer-wise hybrid frameworks}}).
However, due to the sequential nature of model computations, 
such layer-wise partitioning prevents parallel execution between CPU and GPU. 
Furthermore, when workloads of an expert are large, 
computing on the CPU incurs significantly higher 
latency than transferring expert parameters to the GPU and 
processing. 
To address this, 
Fiddler \colorcite{kamahori2024fiddler} and 
HybriMoE \colorcite{zhong2025hybrimoe} 
propose statically assigning individual experts to either CPU or GPU based 
on their workloads (referred to as \textit{\textbf{expert-wise hybrid frameworks}}, 
Figure~\colorref{intro_to_workload}\textcolor{ACMPurple}{b}). 
Experts exceeding a pre-defined workload 
threshold (high-workload experts) are executed on the GPU, while the rest 
(low-workload experts) are handled by the CPU in parallel. However, 
this {static} assignment 
would lead to severe load imbalance between CPU and GPU, 
which results in poor utilization of system resources, thus severely
hindering
the inference performance.

\textbf{Low accuracy in prefetching high-workload experts.} 
Although expert-wise hybrid frameworks enhance inference performance,
the experts assigned to the GPU are determined at runtime and
must be transferred from CPU to GPU before 
computation, imposing significant communication overhead on local PCs 
with limited PCIe bandwidth. For instance, in 
HybriMoE using Mixtral-8$\times$7B,
PCIe transfers account 
for over 60\% of inference time. 
Previous works \colorcite{yi2023edgemoe,yu2025fmoe,eliseev2023fast,zhong2024adapmoe,zhong2025hybrimoe} 
propose prefetching to mitigate the overhead of PCIe transfers.
However, in expert-wise hybrid frameworks, since
GPUs are typically responsible for computing high-workload experts, an
accurate prediction of such experts is essential.
Existing prefetching strategies neglect expert workload characteristics and thus
exhibit poor prediction accuracy on high-workload experts,
resulting in extremely low prefetch accuracy, which 
incurs substantial stall overhead.

\textbf{Inefficient expert-cache design.} 
To further reduce PCIe communication, existing 
methods \colorcite{tang2024hobbit,zhong2024adapmoe,zhong2025hybrimoe} 
employ a portion of the GPU memory as a cache 
for expert parameters. When a cached expert is hit, the corresponding PCIe 
transfer can be avoided. 
For input-dependent and dynamic MoE inference, 
the expert usage varies across tokens.
Therefore, how to design cache replacement strategies is particularly critical.
In expert-wise hybrid frameworks, high-workload experts are 
executed on GPUs, necessitating efficient caching of these experts. Yet, replacement strategies in 
prior works \colorcite{zhong2025hybrimoe,eliseev2023fast} 
neglect workload dynamics, resulting in 
poor cache hit rates. 
For example, HybriMoE achieves only 25.3\% hit rate 
on Mixtral-8$\times$7B, 
significantly limiting cache efficiency.

To tackle the above issues, we holistically redesign the scheduling, prefetching, and caching 
strategies to account for both heterogeneous hardware characteristics of local PCs 
and the dynamic nature of MoE 
workloads—design aspects overlooked by prior work. Specifically, we first formulate the CPU-GPU 
expert assignment problem as a 0-1 integer optimization model to capture heterogeneous execution and 
minimize inference latency. Due to the high overhead of solving this problem directly, we introduce a 
\textbf{Greedy Assignment} strategy that closely approximates the optimal solution with far lower 
computational cost. Second, to enhance prefetch accuracy, we propose a \textbf{Residual-Based 
Prefetching} method that leverages inter-layer residuals to refine features and accurately prefetch 
high-workload experts. Finally, observing the strong temporal correlation of expert workloads across 
tokens, we design a \textbf{Workload-Aware Cache Replacement} strategy 
utilizing the temporal workload information,
significantly improving cache hit rates and inference speed.

We integrate these components into \textbf{DALI}, a workloa\underline{\textbf{D}}-\underline{\textbf{A}}ware
off\underline{\textbf{L}}oad\underline{\textbf{I}}ng MoE framework that substantially accelerates expert-wise 
offloading inference on local PCs.
Experiments across various models and settings
demonstrate 
significant performance improvements. 
Specifically,
DALI on average achieves speedups of 7.62$\times$, 3.80$\times$, 2.45$\times$,
and 2.00$\times$ during the prefill phase, and 3.97$\times$, 2.16$\times$, 1.48$\times$, and 1.32$\times$ 
during decoding compared to the state-of-the-art llama.cpp, 
KTransformers, MoE-Lightning, and HybriMoE, respectively.

\begin{figure}[t]
  \centering
  \vspace{-0.4cm}
  \includegraphics[scale=0.48,trim=0cm 0cm 0cm 0.0cm,clip]{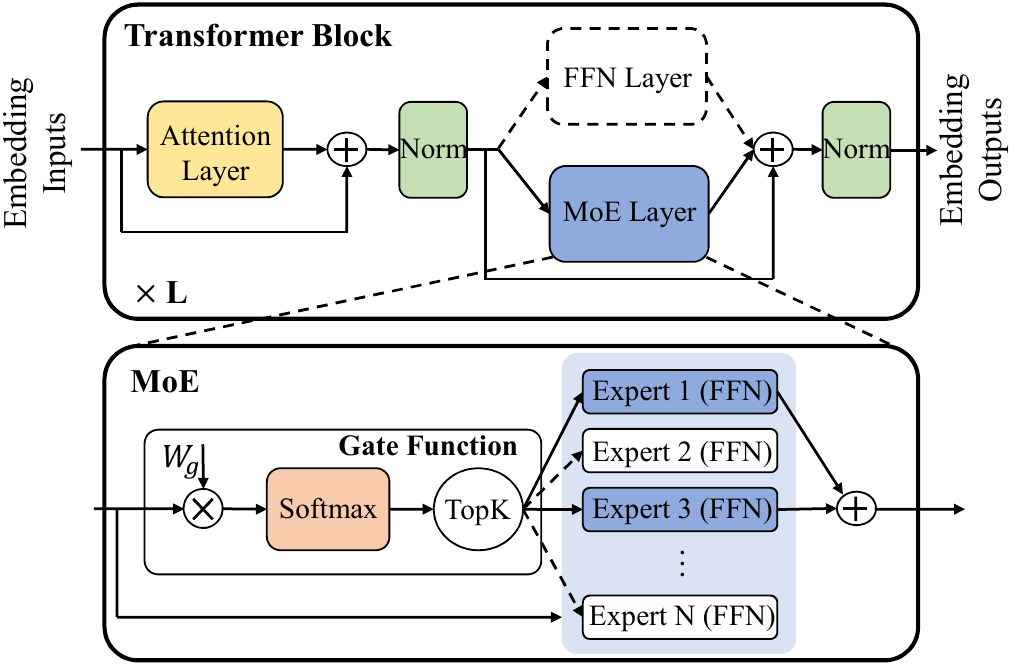}
  \vspace{-0.2cm}
  \caption{An illustration of the MoE architecture.}
  \vspace{-0.6cm}
  \label{moe_background}
\end{figure}

\vspace{-0.2cm}
\section{Background and Related Work}

\subsection{Mixture of Experts (MoE)}
Mixture of Experts (MoE) architectures\colorcite{yuksel2012twenty,jacobs1991adaptive} 
have recently gained widespread adoption 
in LLMs, such as DeepSeek\colorcite{liu2024deepseek,liu2024deepseek_v3,aaa}, 
Mixtral\colorcite{jiang2024mixtral}, Snowflake\colorcite{snowflake}, 
and Qwen\colorcite{chu2024qwen2}.
Traditional LLMs consist 
of multiple stacked transformer blocks\colorcite{vaswani2017attention}, 
whereas MoE architectures replace the 
Feed-Forward Network (FFN) layers within these transformer blocks with MoE layers, 
as illustrated in Figure \colorref{moe_background}. Each MoE layer includes multiple expert sub-networks, 
typically implemented as FFN, and a gate function, which 
dynamically determines which experts should be activated based 
on the input token.
In the gate function, the input ($x\in \mathbb{R}^{1\times d} $) is first multiplied by the 
gate function's weights ($W_g\in \mathbb{R}^{d\times N} $), where $d$ is 
the hidden dimension of the LLM model, and 
$N$ denotes the total number of experts per layer.
Then the resulting values undergo a softmax operation to obtain scores for each expert 
and the top-k highest-scoring experts are selected, where 
$k$ represents the number of activated experts, as the 
following equation:
\begin{equation}
  \label{gate_function}
  \setlength{\abovedisplayskip}{2pt}
  \setlength{\belowdisplayskip}{2pt}
  G(x) = TopK(Softmax(x\cdot W_g)) \ \text{.}
 \end{equation}
Then, each activated expert calculates the corresponding output using $x$. 
Finally, the outputs from these activated 
$k$ experts are combined to generate 
the output of the MoE layer according to the following formula:
\begin{equation}
  \label{moe_output}
  \setlength{\abovedisplayskip}{2pt}
  \setlength{\belowdisplayskip}{2pt}
  MoE_{o}= \sum\limits_{i=1}^k G(x)_i\cdot E(x)_i \ \text{,}
\end{equation}
where $E(x)_i$ denotes the output of the $i$-th selected expert, $G(x)_i$ is the 
activated score of the $i$-th selected expert.

\begin{figure}[t]
  \centering
  \vspace{-0.4cm}
  \includegraphics[scale=0.5,trim=0cm 0cm 0cm 0.0cm,clip]{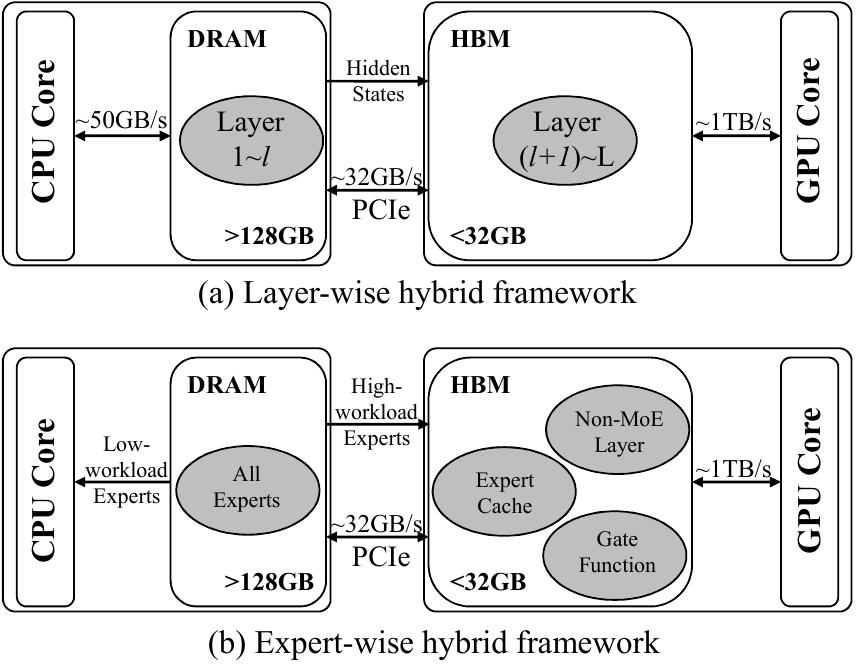}
  \vspace{-0.4cm}
  \caption{The memory hierarchy and the data assignment in hybrid 
  MoE offloading frameworks.}
  \vspace{-0.6cm}
  \label{offload_background}
\end{figure}

\subsection{MoE Offloading}
\label{offload_background_sec}
Offloading frameworks alleviate the storage challenges posed by large-scale MoE models 
by offloading model parameters to secondary storage media such as DRAM, SSD, or HDD, 
thereby mitigating GPU memory pressure without sacrificing model expressiveness. 
Conventional offloading methods\colorcite{rasley2020deepspeed,
he2024fastdecode,jiang2024neo,sheng2023flexgen,
song2024powerinfer,alizadeh2023llm,lee2024infinigen,zhang2024edgeshard}, 
such as DeepSpeed\colorcite{rasley2020deepspeed} 
and FlexGen\colorcite{sheng2023flexgen}, are primarily 
designed for dense LLMs and must fetch all model parameters, 
incurring unnecessary communication overhead when applied to MoE architectures 
due to their inherently sparse activation patterns.

Recently, several offloading frameworks specifically tailored for MoE models have 
been proposed\colorcite{tang2024hobbit,hwang2024pre,zhong2024adapmoe,yu2025fmoe,
eliseev2023fast,zhong2025hybrimoe,yi2023edgemoe,llama.cpp,ktransformers,
kamahori2024fiddler,cao2025moe}. 
Although these frameworks introduce innovative
prefetching techniques and caching strategies, they still require experts to be 
transferred to GPUs before execution. Due to PCIe bandwidth limitations and the 
substantial parameter counts of MoE, transferring expert 
parameters to GPUs introduces considerable latency overhead.

To mitigate this problem, recent studies\colorcite{llama.cpp,ktransformers,
kamahori2024fiddler,zhong2025hybrimoe}
propose hybrid MoE offloading systems, effectively leveraging 
CPU computational resources for offloaded expert-related computations, thereby significantly 
reducing PCIe transfers and enhancing inference speed. Nevertheless, these existing 
solutions exhibit several critical shortcomings. 
As shown in Figure~\colorref{offload_background}\textcolor{ACMPurple}{a},
layer-wise hybrid frameworks like Ktransformers\colorcite{chen2025ktransformers}
and llama.cpp\colorcite{llama.cpp} 
assign assign the first $l$ layers to the CPU and layers $l$ through $L$ to the GPU,
executing each MoE layer exclusively on one device.
However, the layers on the CPU would incur significant inference 
latency when expert workloads become large, due to inherently slower CPU performance.
Expert-wise hybrid frameworks such as Fiddler\colorcite{kamahori2024fiddler} 
and HybriMoE\colorcite{zhong2025hybrimoe} mitigate this problem by
allocating experts based
on workload size, as shown in Figure~\colorref{offload_background}{b}. 
However, such static allocation 
strategies neglect real-time CPU-GPU load imbalances and 
result in inefficient utilization of heterogeneous computational resources, 
ultimately increasing inference latency. 
{MoE\mbox{-}Lightning~\colorcite{cao2025moe} adopts a performance\mbox{-}analysis model to
offline\mbox{-}search deployment strategies, but it insufficiently models MoE
characteristics and thus performs poorly.}

\section{Motivation}
\label{motivation_sec}

\subsection{The Necessity of Dynamic Expert Assignment}
\label{motivation_sec_1}
Due to PCIe bandwidth limitations and the large number of MoE expert parameters, 
an increasing number of MoE offloading frameworks adopt hybrid CPU-GPU execution to 
accelerate inference\colorcite{llama.cpp,zhong2025hybrimoe,
ktransformers,kamahori2024fiddler}. 
Thanks to assign experts instead of layers to either CPU or GPU,
\textit{\textbf{expert-wise hybrid frameworks}}\colorcite{zhong2025hybrimoe,kamahori2024fiddler} 
can perform high-workload expert computation on GPU and
achieve better performance than 
\textit{\textbf{layer-wise hybrid frameworks}}\colorcite{llama.cpp,ktransformers}.
However, existing expert-wise works\colorcite{zhong2025hybrimoe,kamahori2024fiddler} 
assign experts purely based on workload, i.e., low-workload experts 
are executed on the CPU, while high-workload experts are transferred to the GPU for 
computation. This manner
overlooks the 
parallel characteristics of heterogeneous systems and
introduces serious 
CPU-GPU load imbalance. 
As shown in Figure \colorref{workload_imb_motivation}, 
we measure the execution time of CPU- and GPU-assigned experts 
under the assignment policy of Fiddler\colorcite{kamahori2024fiddler} 
on 
DeepSeek-V2-Lite and Qwen-1.5 across different batch sizes. 
The significant gap between CPU and GPU execution time indicates 
severe imbalance. 
{When the batch size is small, the experts' workloads are often light, 
so the static assignment allocates 
most experts to the CPU, leading to much longer CPU execution time.
The GPU idles waiting for the CPU, yielding low GPU utilization.
As batch size grows, more
high\mbox{-}workload experts emerge, and the imbalance reverses.
}

\begin{figure}[t]
  \centering
  \vspace{-0.4cm}
  \includegraphics[scale=0.37,trim=0cm 0cm 0cm 0.0cm,clip]{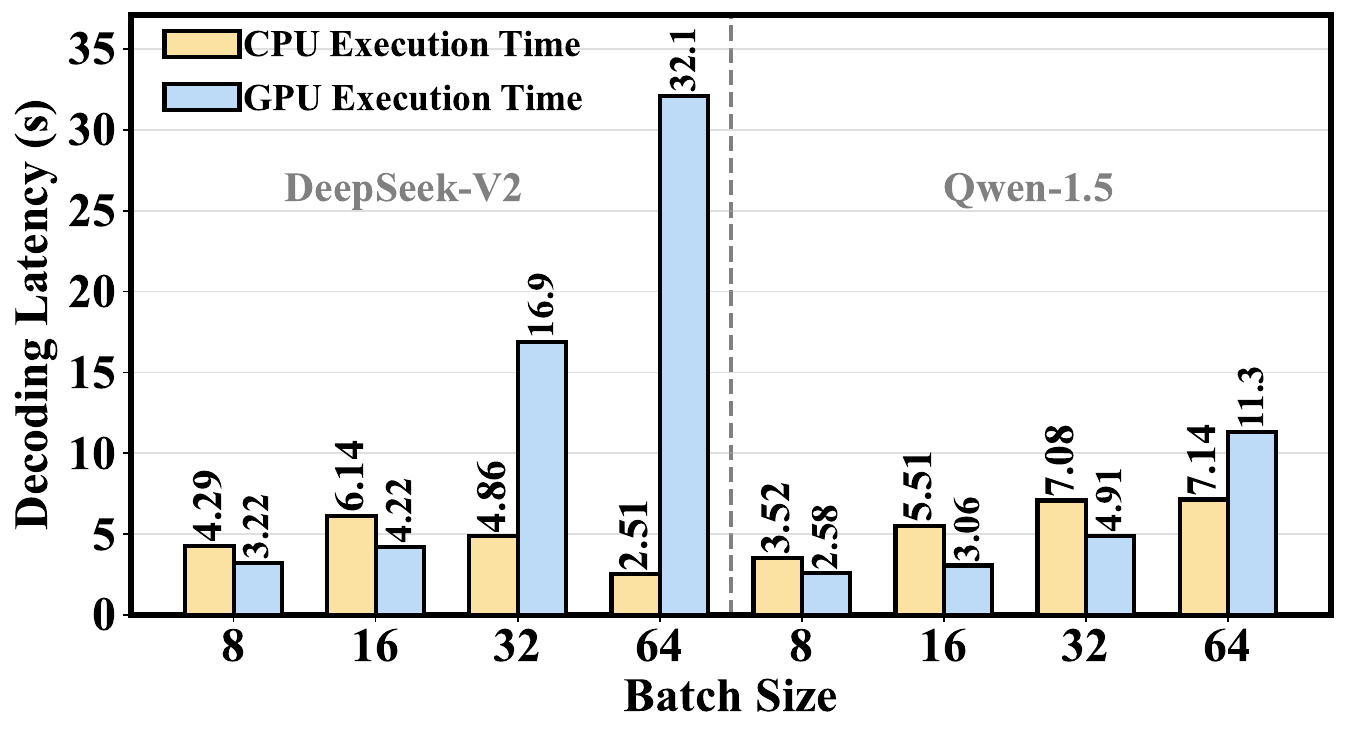}
  \vspace{-0.8cm}
  \caption{Comparison of execution time between CPU- and GPU-assigned experts 
  under different batch sizes. 
  The prefill and decoding length are both 32.}
  \vspace{-0.3cm}
  \label{workload_imb_motivation}
\end{figure}

To address this, we propose a workload-aware \textbf{Greedy Assignment} strategy.
We first formulate expert assignment as a 0-1 integer optimization problem that 
explicitly captures system parallelism to derive an optimal expert execution
schedule between CPU and GPU.
To mitigate the high latency overhead of precise solving, 
we further introduce a heuristic greedy strategy 
that achieves near-optimal assignment with significantly lower solving latency.

\begin{figure}[t]
  \centering
  \vspace{-0.2cm}
  \includegraphics[scale=0.36,trim=0cm 0cm 0cm 0.0cm,clip]{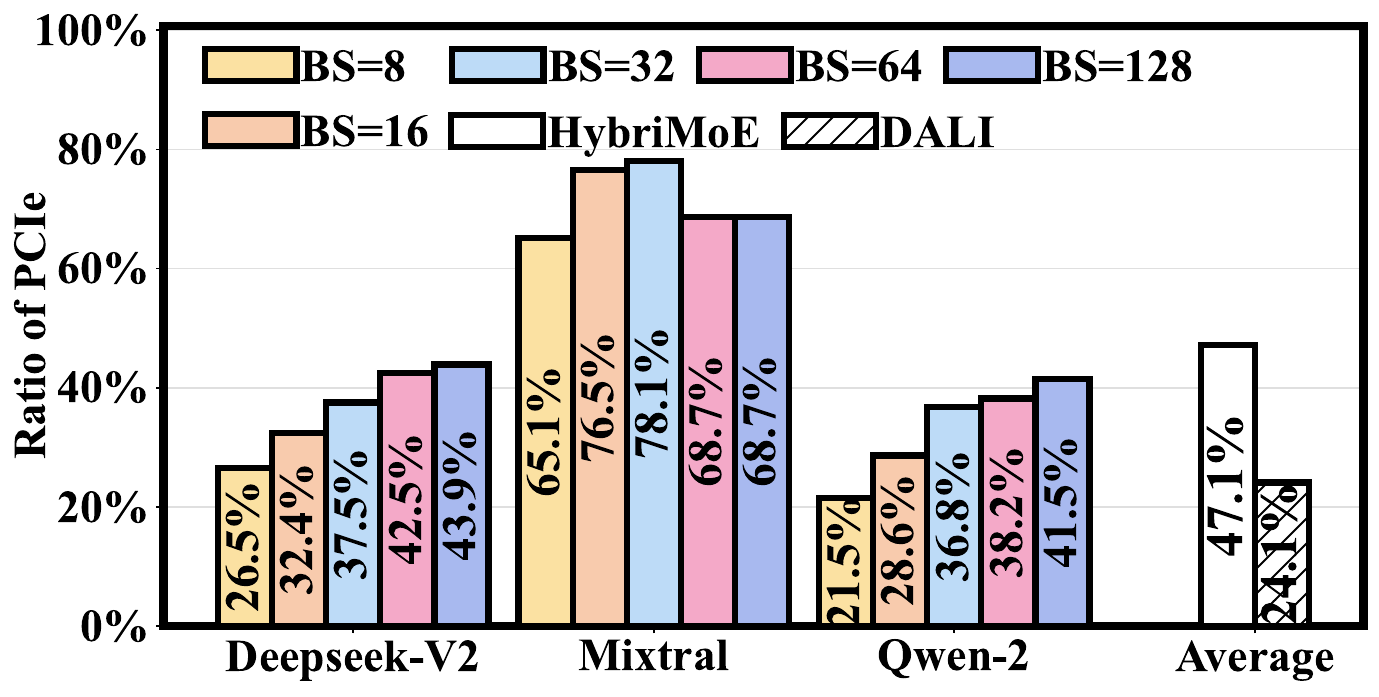}
  \vspace{-0.7cm}
  \caption{Proportion of PCIe transfer time relative to total inference time under various batch sizes. 
  The prefill and decoding length are both 32. `BS' denotes the batch size. 
  {As shown in 
  ``Average'', with our proposed techniques, DALI significantly reduces PCIe traffic compared
  against HybriMoE.}}
  \vspace{-0.6cm}
  \label{pcie_motivation}
\end{figure}

\subsection{The Importance of Optimizing Prefetching}
\vspace{-0.1cm}
\label{sec:prefetch_motivation}
Although 
expert-wise hybrid frameworks significantly improve inference efficiency, especially with our 
Greedy Assignment strategy,
PCIe transfer of the experts assigned to the GPU still remains a dominant performance bottleneck. As shown 
in Figure~\colorref{pcie_motivation}, PCIe transfer accounts for up 
to 78.1\% of total execution time under 
hybrid execution, underscoring the urgency to reduce communication overhead.

Prefetching is a widely adopted technique to overlap computation and communication, 
effectively hiding PCIe latency
in series 
of MoE offloading frameworks\colorcite{fang2025accurate,hwang2024pre,xue2024moe,zhong2025hybrimoe,
zhong2024adapmoe,eliseev2023fast,cai2024read,du2024sida,yu2025fmoe}. 
However, unlike prior frameworks that need all activated experts on the GPU, 
hybrid MoE framework only assigns high-workload experts to the GPU. 
This places stricter demands on prefetching: it must accurately prefetch 
experts with large workloads.
As shown in Table~\colorref{prefetch_acc_motivation}, when 
adopting the prefetching method in statistical-based
EdgeMoE\colorcite{yi2023edgemoe} and feature-based HybriMoE\colorcite{zhong2025hybrimoe} to 
prefetch high-workload experts, the prefetch mechanism shows 
poor accuracy. 
Much worse, as illustrated in Figure \colorref{prefetch_motivation}, 
low prediction 
accuracy results in minimal performance gains from prefetching 
{due to the necessary re-fetch 
when prefetch misses}, motivating the need 
for a more accurate strategy tailored to high-workload expert prediction in hybrid frameworks.

To this end, inspired by the residual between layers, we 
propose a \textbf{Residual-Based Prefetching} strategy. It uses the residual 
information between adjacent MoE layers to 
improve the precision of prefetching high-workload experts, 
thus
unleashing acceleration benefits from prefetching.

\begingroup
\setlength{\tabcolsep}{3.5pt}   
\renewcommand{\arraystretch}{0.95}

\begingroup
\setlength{\tabcolsep}{2.5pt}
\renewcommand{\arraystretch}{0.85} 
\setlength{\extrarowheight}{-0.6pt} 
\setlength{\aboverulesep}{0.4ex}
\setlength{\belowrulesep}{0.4ex}
\begin{table}[t]
\vspace{-0.5cm}
\caption{{Prefetch accuracy for predicting experts with different workload levels.
  Topk=$k$ indicates prediction of the top $k$ most high-workload experts.}}
  \vspace{-0.2cm}
  \label{prefetch_acc_motivation}
  \begin{center}
    \begin{tabular}{@{} c l c c c c c @{}} 
      \toprule[1.5pt]
      \multicolumn{2}{c}{} &
      \multirow{2}{*}{\textbf{Method}} &
      \multicolumn{4}{c}{\textbf{Batch size}} \\
      \cmidrule(lr){4-7}
      \multicolumn{2}{c}{} &  &
      \textbf{8} & \textbf{16} & \textbf{32} & \textbf{64} \\
      \midrule[1pt]

      \multirow{4}{*}{\shortstack[c]{DeepSeek-\\V2-Lite}}
        & \multirow{2}{*}{Topk=1} & EdgeMoE  & 35.3\% & 24.5\% & 14.6\% & 11.8\% \\
        &                         & HybriMoE & 36.7\% & 32.7\% & 35.8\% & 40.6\% \\\cmidrule(lr){2-7}
        & \multirow{2}{*}{Topk=2} & EdgeMoE  & 34.7\% & 25.0\% & 20.1\% & 15.4\% \\
        &                         & HybriMoE & 45.0\% & 40.1\% & 39.3\% & 38.9\% \\
      \midrule[1pt]

      \multirow{4}{*}{\shortstack[c]{Mixtral-\\8x7B}}
        & \multirow{2}{*}{Topk=1} & EdgeMoE  & 25.7\% & 25.4\% & 28.6\% & 30.9\% \\
        &                         & HybriMoE & 51.1\% & 53.4\% & 48.4\% & 46.8\% \\\cmidrule(lr){2-7}
        & \multirow{2}{*}{Topk=2} & EdgeMoE  & 40.3\% & 40.1\% & 44.7\% & 48.5\% \\
        &                         & HybriMoE & 65.2\% & 63.7\% & 58.9\% & 56.4\% \\
      \bottomrule[1.5pt]
    \end{tabular}
  \end{center}
  \vspace{-0.5cm}
\end{table}
\endgroup

\begin{figure}[t]
  \centering
  \includegraphics[scale=0.36,trim=0cm 0cm 0cm 0.0cm,clip]{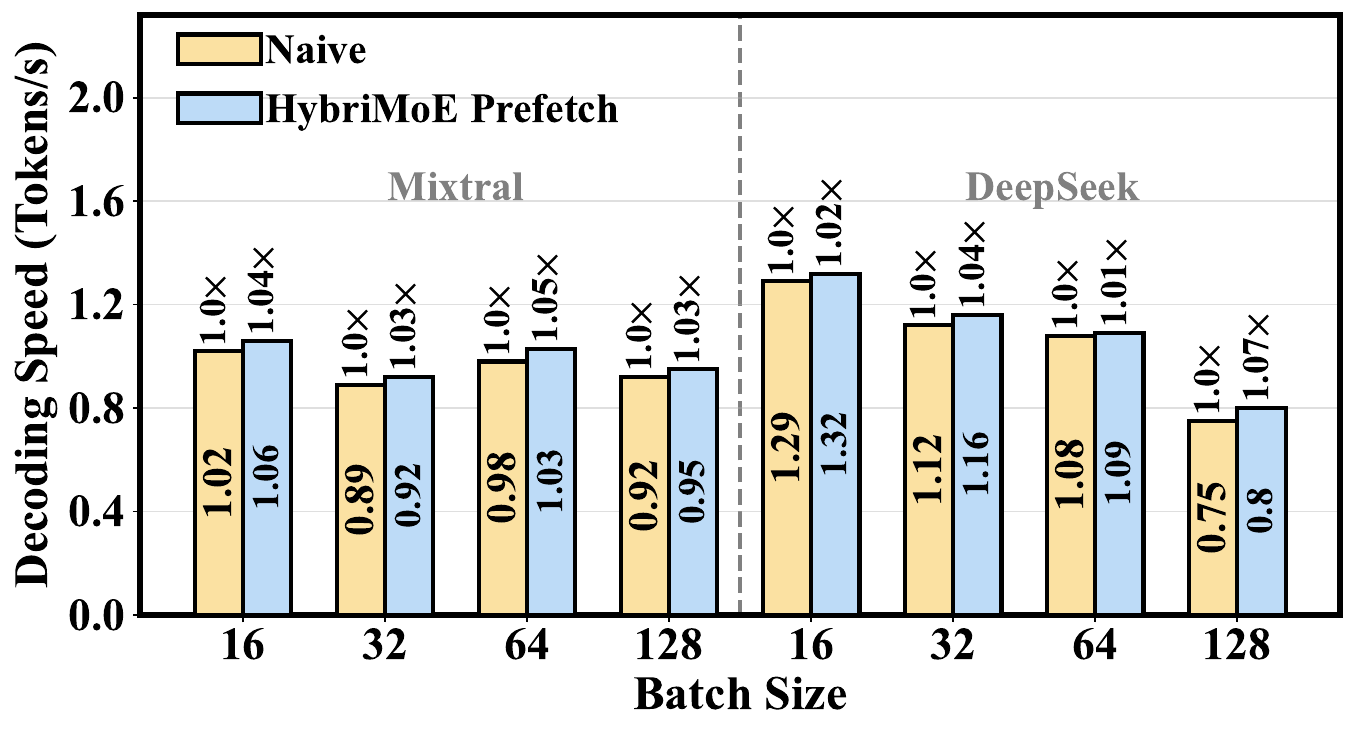}
  \vspace{-0.4cm}
  \caption{Speedup achieved by HybriMoE's prefetching strategy compared to no 
  prefetching under different batch sizes on DeepSeek-V2-Lite and Mixtral-8x7B.}
  \vspace{-0.4cm}
  \label{prefetch_motivation}
\end{figure}

\subsection{Challenge in Cache Utilization}
\label{sec:cache_motivation}

To better utilize the limited GPU memory, many prior MoE offloading 
frameworks maintain an expert cache on the GPU to store a subset of 
experts\colorcite{zhong2025hybrimoe,xue2024moe,zhong2024adapmoe}. 
Upon the expert cache hit, the
data traffic between CPU and GPU can be avoided.
However, due to the input-dependent expert activation, 
the set of experts used during inference varies dynamically. As a result,
cache replacement policies are crucial to maintaining high cache hit rates.
Existing methods, such as FastMoE\colorcite{eliseev2023fast} adopts 
traditional LRU policies, while HybriMoE\colorcite{zhong2025hybrimoe} uses
activation scores of experts to update the cache. However, in expert-wise MoE offloading frameworks, 
GPUs primarily 
compute high-workload experts, which implies that cached experts should preferentially be 
high-workload ones. Unfortunately, as shown in Figure \colorref{cache_motivation_hit_rate}, 
{neither LRU nor score-based strategies (e.g., HybriMoE) consider expert workload, 
resulting in poor cache hit rates,}
which severely limits the benefits from 
caching. 

\begin{figure}[t]
  \centering
  \vspace{-0.4cm}
  \includegraphics[scale=0.36,trim=0cm 0cm 0cm 0.0cm,clip]{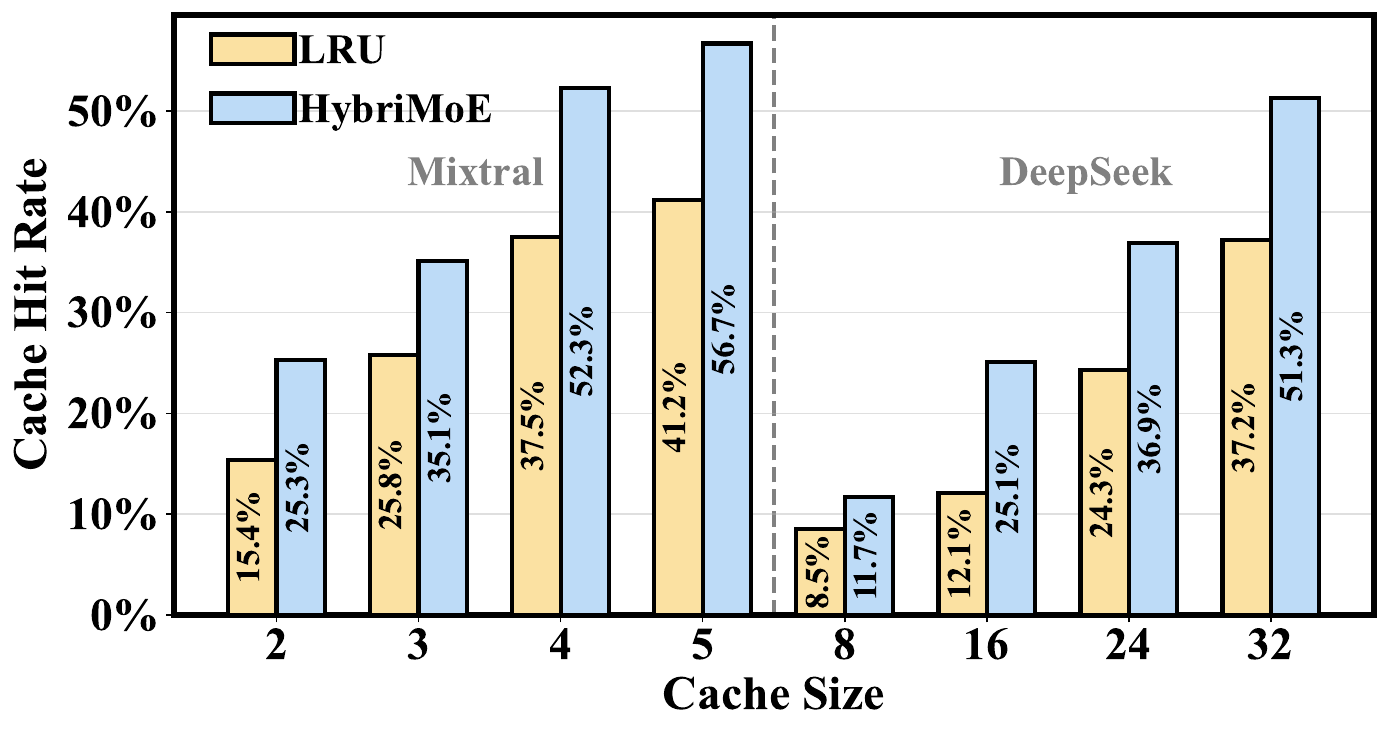}
  \vspace{-0.8cm}
  \caption{Cache hit rates of LRU and HybriMoE's replacement strategies under different cache sizes on 
  DeepSeek-V2-Lite and Mixtral-8x7B. 
  Cache size is measured in the number of cached experts in each MoE layer.
  }
  \vspace{-0.2cm}
  \label{cache_motivation_hit_rate}
\end{figure}

\begin{figure}[t]
  \centering
  \vspace{-0.2cm}
  \includegraphics[scale=0.36,trim=0cm 0cm 0cm 0.0cm,clip]{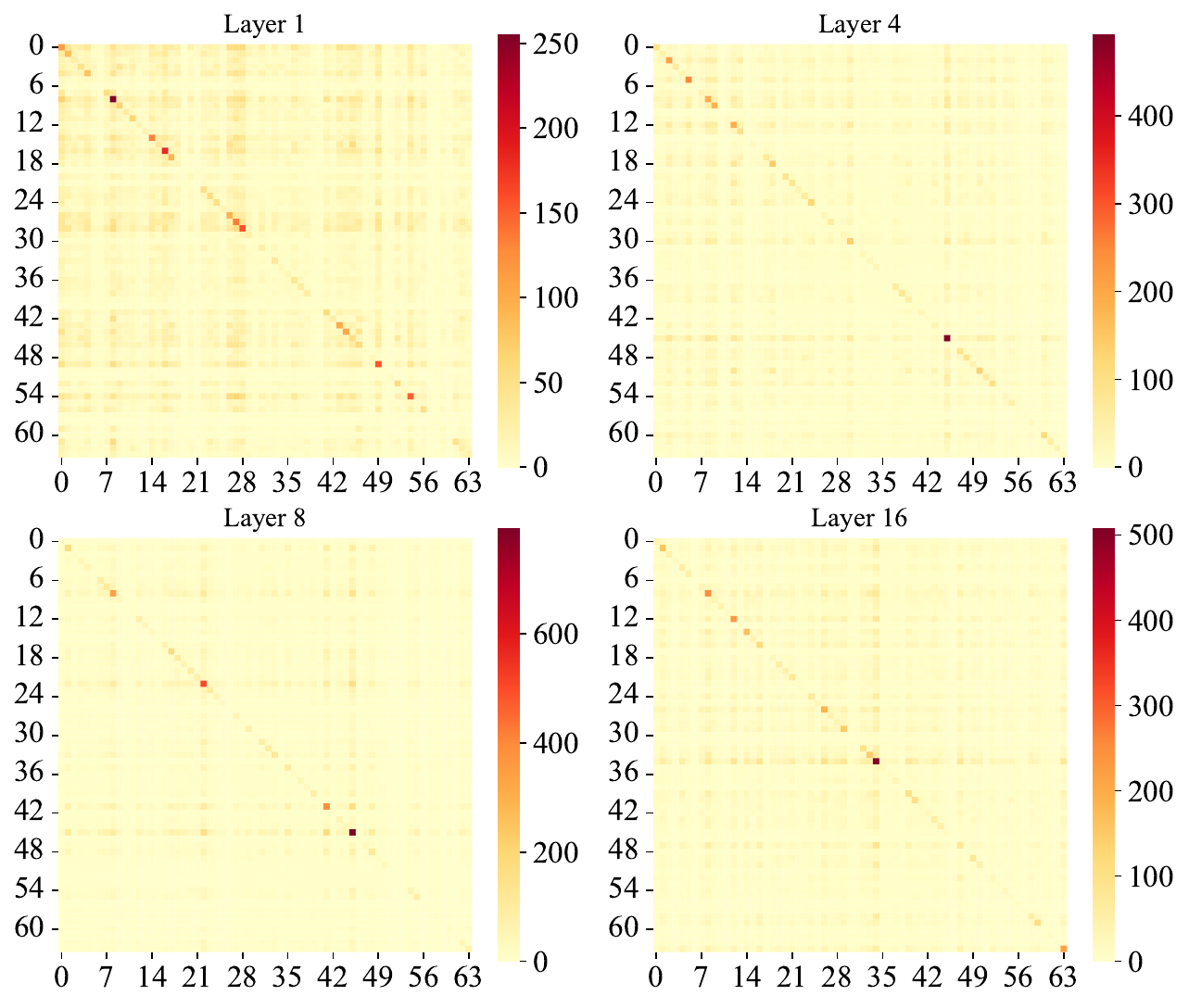}
  \vspace{-0.2cm}
  \caption{Correlation of high-workload expert activation between adjacent 
  tokens in layers 1, 4, 8, and 16 of Mixtral-8$\times$7B. The x- and y-axes 
  represent high-workload experts (top 3 by workload) activated for adjacent tokens.}
  \vspace{-0.4cm}
  \label{workload_aware_motivation}
\end{figure}

MoE experts are typically responsible for different knowledge domains, 
and adjacent tokens within a sequence often share similar semantics\colorcite{zhu2025mata,fedus2022switch,zhou2022mixture}. 
Motivated by this, we investigate the temporal locality of high-workload expert usage—specifically, whether 
experts with high workload at token $i$ tend to remain high-workload at token $i+1$.
Figure \colorref{workload_aware_motivation} presents 
a heatmap where each cell at position $(m, n)$ 
records the frequency with which expert $m$ is a 
high-workload expert at token $i$ and expert $n$ is high-workload at token $i{+}1$.
The pronounced diagonal pattern in the heatmap indicates that if 
an expert is high-workload for token $i$, 
it is highly likely to remain so for token $i+1$.

Based on this observation, we design a \textbf{Workload-Aware Cache 
Replacement} strategy, which 
updates the expert cache according to the workload history of previous tokens, significantly improving 
cache hit rates 
and leading to further acceleration of MoE offloading inference.

\section{Design}
Figure~\colorref{overview_fig}
illustrates the overview 
of our DALI framework. During deployment, 
all expert weights are stored in CPU DRAM. 
Additionally, for each MoE layer, we randomly select a 
fixed number of experts (defined as ${cache\_size}$) 
to be cached in GPU memory as an expert cache.
When performing the MoE layer computation, 
DALI first determines the expert assignment 
across CPU and GPU using our proposed {Greedy Assignment} strategy at runtime, 
based on the current layer's expert activation pattern. 
The activated experts are then processed in parallel by CPU and GPU. 
Meanwhile, a separate work stream is launched 
to execute our {Residual-Based Prefetching} strategy, 
which predicts and prefetches high-workload experts required for the next MoE layer.
If an expert assigned to the GPU is already cached in GPU memory, 
its parameters are directly used for computation ($E_1$). Otherwise, 
the expert weights are loaded from DRAM to GPU ($E_3,E_7$). 
DALI
updates the expert cache using our {Workload-Aware Cache Replacement} policy 
to maintain high cache hit rates.
The above process is repeated iteratively until the inference is complete.

\vspace{-0.1cm}
\subsection{Greedy Assignment Strategy}
\vspace{-0.1cm}

To achieve load balance between the CPU and GPU, and identify the optimal 
expert assignment schedule for reducing the inference latency, 
we propose a Greedy Assignment strategy exploiting the 
heterogeneous hardware characteristics and workload 
properties.

To minimize the execution time of the MoE layer, we first formulate the following optimization objective:
\begin{equation}
  \setlength{\abovedisplayskip}{2pt}
  \setlength{\belowdisplayskip}{2pt}
\min \max(T_{\text{gpu}}, T_{\text{cpu}}),
\label{eq:objective}
\end{equation}
where $T_{\text{gpu}}$ and $T_{\text{cpu}}$ represent the total 
execution times of experts assigned to 
the GPU and CPU, respectively. Due to the parallelism of heterogeneous 
systems, the MoE layer's latency is determined by the slower of the two devices.

\begin{figure}[t]
  \centering
  \vspace{-0.4cm}
  \includegraphics[scale=0.45,trim=0cm 0cm 0cm 0.0cm,clip]{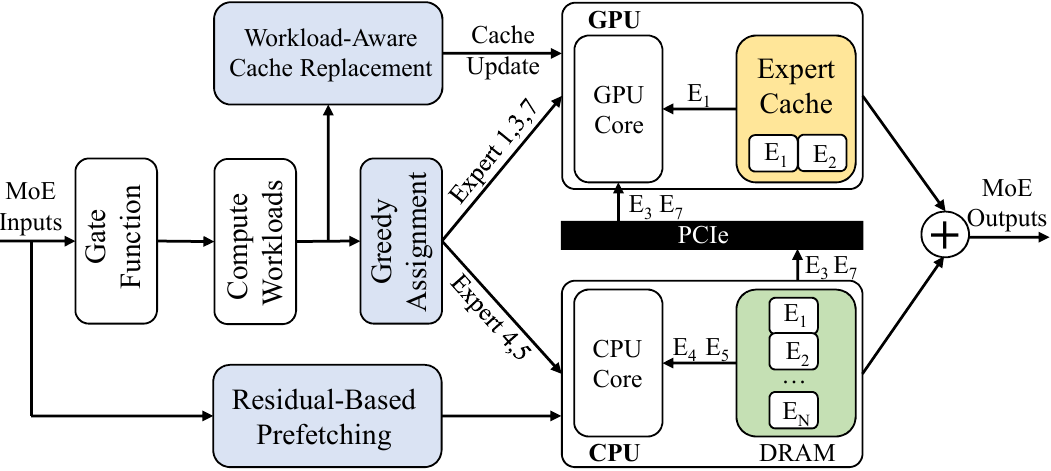}
  \caption{The overview of our DALI framework.}
  \vspace{-0.5cm}
  \label{overview_fig}
\end{figure}

To obtain the optimal assignment schedule, 
We use two binary vectors, $C, G \in \{0,1\}^N$, 
to denote the assignment of $N$ routed experts: 
$C_i = 1$ indicates that expert $i$ is assigned to the CPU, 
while $G_i = 1$ denotes assignment to the GPU.
The CPU execution time is calculated as:
\begin{equation}
  \setlength{\abovedisplayskip}{2pt}
\setlength{\belowdisplayskip}{2pt}
    T_{\text{cpu}} = \sum_{i=1}^{N}t_{\text{cpu}}(w_i)\cdot C_i \ \text{,}
    \label{eq:cpu_time}
\end{equation}
where $w_i$ is the workload of expert $i$, and $t_{\text{cpu}}(w_i)$ gets the 
execution time of expert $i$ on the CPU for the given workload.
The GPU execution time is calculated as:
\begin{equation}
  \setlength{\abovedisplayskip}{1pt}
\setlength{\belowdisplayskip}{1pt}
  T_{\text{gpu}} = \sum_{i=1}^{N} t_{\text{gpu}}(w_i) \cdot G_i \ \text{,}
  \label{eq:gpu_time}
\end{equation}
where $t_{\text{gpu}}(w_i) = \max(\text{Trans}_{\text{expert}}(w_i), \text{compute}_{\text{expert}}(w_i))$,
$\text{Trans}_{\text{expert}}(w_i)$ denotes the time for PCIe transmission of 
expert $i$, and $\text{compute}_{\text{expert}}(w_i)$ denotes the GPU execution time 
of expert $i$. 
Due to pipeline parallelism, the GPU execution time for an expert is the maximum of these two.
The $\text{Trans}_{\text{expert}}(w_i)$ is defined as:
\begin{equation}
  \setlength{\abovedisplayskip}{1pt}
\setlength{\belowdisplayskip}{1pt}
  \text{Trans}_{\text{expert}}(w_i) =
  \begin{cases}
  0, & w_i = 0 \\
  trans\_time, & w_i > 0
  \end{cases} \text{,}
  \label{eq:expert_transfer_cost}
\end{equation}
where $trans\_time$ 
represents the time required to transfer the 
weights of a single expert from DRAM to GPU via PCIe. 
All hardware-specific timing values (e.g., $trans\_time$, $t_{\text{cpu}}(w_i)$) can be obtained through warm-up profiling before 
execution and can be reused for later inference.

In addition to timing considerations, the optimization of Equation~\eqref{eq:objective} 
is subject to the following constraints:
\\
\textbf{1. Expert activation constraint:}
\begin{equation}
  \setlength{\abovedisplayskip}{2pt}
\setlength{\belowdisplayskip}{2pt}
  \sum_{i=1}^{N}(C_i + G_i) = expert\_num \ \text{,}
\end{equation}
where $expert\_num$ is the total number of activated experts in the MoE layer
for given inputs.
\\
\textbf{2. Mutual exclusion constraint:}
    \begin{equation}
      \setlength{\abovedisplayskip}{3pt}
\setlength{\belowdisplayskip}{3pt}
       0 \leq C_i + G_i \leq 1, \quad \forall i = 1, 2, \dots, N \ \text{.}
    \end{equation}
This constraint indicates that each expert can only be assigned to either the CPU or 
the GPU, not both. When $C_i = G_i = 0$, it means expert $i$ is not activated.
\\
\textbf{3. GPU memory constraint:}
\setlength{\abovedisplayskip}{3pt}
\setlength{\belowdisplayskip}{3pt}
    \begin{equation}
        \sum_{i=1}^{N} G_i \cdot \text{size}(E_i) \leq M_{\text{gpu}} \ \text{,}
    \end{equation}
where $\text{size}(E_i)$ denotes the memory required by expert $i$, and this 
constraint ensures that the total memory consumption on the GPU does not exceed 
its memory capacity $M_{\text{gpu}}$.

\begingroup
\tighttextfloatsepOnce{4pt plus 0pt minus 0pt}
\begin{algorithm}[t]
  \caption{Greedy Assignment Strategy}
  \label{alg:greedy}
  \begin{algorithmic}[1]
  \State $C = [0, \dots, 0]_N$, $G = [0, \dots, 0]_N$
  \State $T_{\text{cpu}} = 0$, $T_{\text{gpu}} = 0$
  \State $t_{\text{gpu}} = [t_{\text{gpu}}(w_1), \dots, t_{\text{gpu}}(w_N)]$
  \State $t_{\text{cpu}} = [t_{\text{cpu}}(w_1), \dots, t_{\text{cpu}}(w_N)]$
  \State $\text{sorted\_indices} = \text{argsort}(|t_{\text{gpu}} - t_{\text{cpu}}|, descending=\text{True})$
  \ForAll{$\text{idx} \in \text{sorted\_indices}$}
      \State $gpu\_time = t_{\text{gpu}}[\text{idx}]$
      \State $cpu\_time = t_{\text{cpu}}[\text{idx}]$
      \If{$gpu\_time==0 \ \&\& \ cpu\_time==0$}
          \State \textbf{continue}
      \EndIf
      \If{$T_{\text{gpu}} + gpu\_time \leq T_{\text{cpu}} + cpu\_time$}
          \State $G[\text{idx}] = 1$
          \State $T_{\text{gpu}} = T_{\text{gpu}} + gpu\_time$
      \Else
          \State $C[\text{idx}] = 1$
          \State $T_{\text{cpu}} = T_{\text{cpu}} + cpu\_time$
      \EndIf
  \EndFor
  \State \Return $C, G$
  \end{algorithmic}
\end{algorithm}
\endgroup

By solving the optimization problem in Equation~\eqref{eq:objective} under these constraints, 
we obtain the optimal assignment of activated experts, which greatly improves the computation efficiency.
However, the precise solving process introduces significant latency overhead and diminishes
the performance gain. Therefore, we propose a heuristic Greedy Assignment strategy 
that approximates the optimal solution with minimal solving cost,
thereby further accelerating MoE inference.

Algorithm~\colorref{alg:greedy} illustrates our Greedy Assignment strategy. The key idea of the 
greedy strategy is to prioritize assigning experts whose CPU and 
GPU execution times differ most significantly to reduce overall inference latency.
First, we initialize the 
assignment variables $C$ and $G$, as well as the total 
execution times $T_{\text{gpu}}$ and $T_{\text{cpu}}$. Next, based on the expert activation of 
each token, we obtain the workload of each expert and compute its expected execution 
time on both CPU and GPU 
using $t_{cpu}(w_i)$ and $t_{gpu}(w_i)$, as shown in lines 1-4.
In line 5, we 
sort the experts in descending order based on the absolute difference between their 
CPU and GPU execution times. 
The algorithm then iterates over this list: 
for each expert, 
if assigning it to the GPU results 
in a lower cumulative latency than assigning it to the CPU, 
it is allocated to the GPU (lines 12-14); otherwise, it is assigned to the CPU (lines 15-17).
Moreover, if an expert is not activated, we do not assign it (lines 9-10).

After all experts have been processed, the Greedy Assignment strategy yields the final 
assignment 
vectors $C$ and $G$, which will be used in the subsequent expert computation. Experiments demonstrate 
that the assignment schedule produced by the greedy strategy 
achieves up to 92\% of the performance of the 
optimal solution, while incurring only 5\% (v.s. 55\% of the optimal solution)
end-to-end latency overhead. This enhances the 
acceleration benefit brought by dynamic assignment and 
further improves the inference performance.

\subsection{Residual-Based Prefetching}

As analyzed in Section~\colorref{sec:prefetch_motivation}, 
previous approaches suffer from extremely low accuracy in prefetching
high-workload experts, which severely limits the potential 
acceleration benefit from prefetching. 
To address this, we propose a Residual-Based Prefetching method that 
leverages the residual between adjacent MoE layer features to adjust the 
features used for prefetching. 
This significantly improves the accuracy of the high-workload expert 
prefetching, thereby improving the inference speedup.

Figure~\colorref{res_prefetch} presents our Residual-Based 
Prefetching method. 
Firstly, our approach adopts the 
feature-based prefetching scheme: it uses the input features of the current MoE layer's 
gate function and the gate function of the next MoE layer to predict which experts
will be activated in the next layer.
However, as our analyses indicate, naïvely using only the raw input features results in very low accuracy when 
predicting high-workload experts. Therefore, 
inspired by the similarity between inputs of adjacent MoE 
gate functions due to residual connections, our Residual-Based Prefetching strategy further applies a 
residual correction to the current input features. This makes them better approximate 
the next MoE layer's gating input (measured by the 
cosine similarity analyzed in Appendix~\colorref{prefetch_analysis_appendix}) and thus improves 
the accuracy of expert activation prediction. The transformation is defined as:
\begin{equation}
\setlength{\abovedisplayskip}{3pt}
  \begin{aligned}
    \tilde{h}^{(l)} = \text{hidden\_states}^{(l)} + {res\_vec}^{(l)} \text{,} \\
    {predict\_expert}^{(l+1)} = \text{gate\_func}^{(l+1)}(\tilde{h}^{(l)}) \text{,}
  \end{aligned}
  \label{eq:residual_prefetch}
\setlength{\belowdisplayskip}{3pt}
\end{equation}
where $\text{hidden\_states}^{(l)}$ is the 
input to the $l$-th MoE gate, ${res\_vec}^{(l)}$ is the layer-specific 
residual vector of layer $l$, 
and $\text{gate\_func}^{(l+1)}$ is the gate function for layer $l+1$. 
The ${res\_vec}^{(l)}$ has the same dimension with the $\text{hidden\_states}^{(l)}$ 
along the feature axis,
and is shared across all tokens.
Once the predicted 
activated experts of the next layer obtained, we count the number of tokens routed to each expert, and
the top-$k$ high-workload experts are selected for prefetching.

\begin{figure}[t]
  \centering
  \vspace{-0.4cm}
  \includegraphics[scale=0.48,trim=0cm 0cm 0cm 0.0cm,clip]{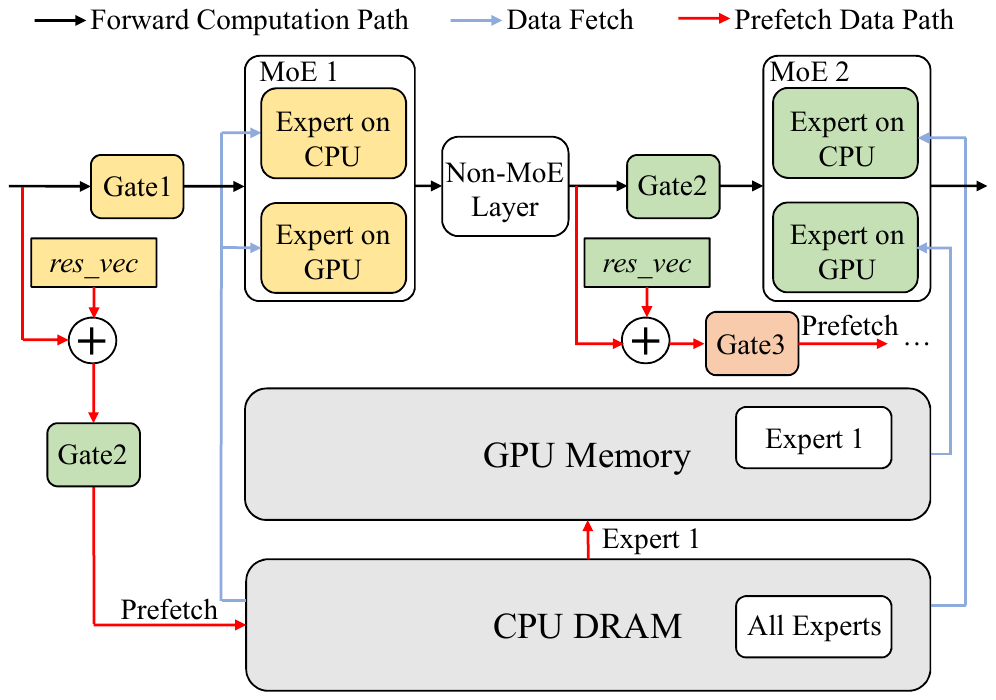}
  \vspace{-0.2cm}
  \caption{Illustration of our Residual-Based Prefetching method.}
  \vspace{-0.4cm}
  \label{res_prefetch}
\end{figure}

Note that obtaining the residual vector requires no fine-tuning or retraining.
It can be constructed offline by running inference on a small
calibration dataset (described in Section~\colorref{experimental_setup}) and can be reused on 
various downstream tasks. 
The residual 
vector for layer $l$ is computed as:
\begin{equation}
\setlength{\abovedisplayskip}{3pt}
\setlength{\belowdisplayskip}{3pt}
{res\_vec}^{(l)} = \frac{1}{N} \sum_{i=1}^{N} \left( \text{hidden\_states}^{(l+1)}_i - 
\text{hidden\_states}^{(l)}_i \right) \ \text{,}
\label{eq:residual_matrix}
\end{equation}
where $N$ is the number of tokens in the calibration dataset. 
Each MoE layer maintains its own residual vector,
except for the last one, which does not require prefetching
for any subsequent layer.

\subsection{Workload-Aware Cache Replacement}
\label{cache_sec}

As analyzed in Section~\colorref{sec:cache_motivation}, 
prior works ignore the influence of dynamic workloads when 
designing cache replacement strategies for the expert cache on the GPU, 
resulting in low cache hit rates. 
Fortunately,
our analysis reveals a strong correlation in high-workload 
expert activations between adjacent tokens. Motivated by this observation, we propose 
a {{Workload-Aware Cache Replacement}} strategy that updates the 
expert cache based on dynamic workload patterns. This 
approach significantly improves the cache hit rate and, in turn, accelerates 
the inference performance of our DALI framework.

\begin{figure}[t]
  \centering
  \vspace{-0.4cm}
  \includegraphics[scale=0.48,trim=0cm 0cm 0cm 0.0cm,clip]{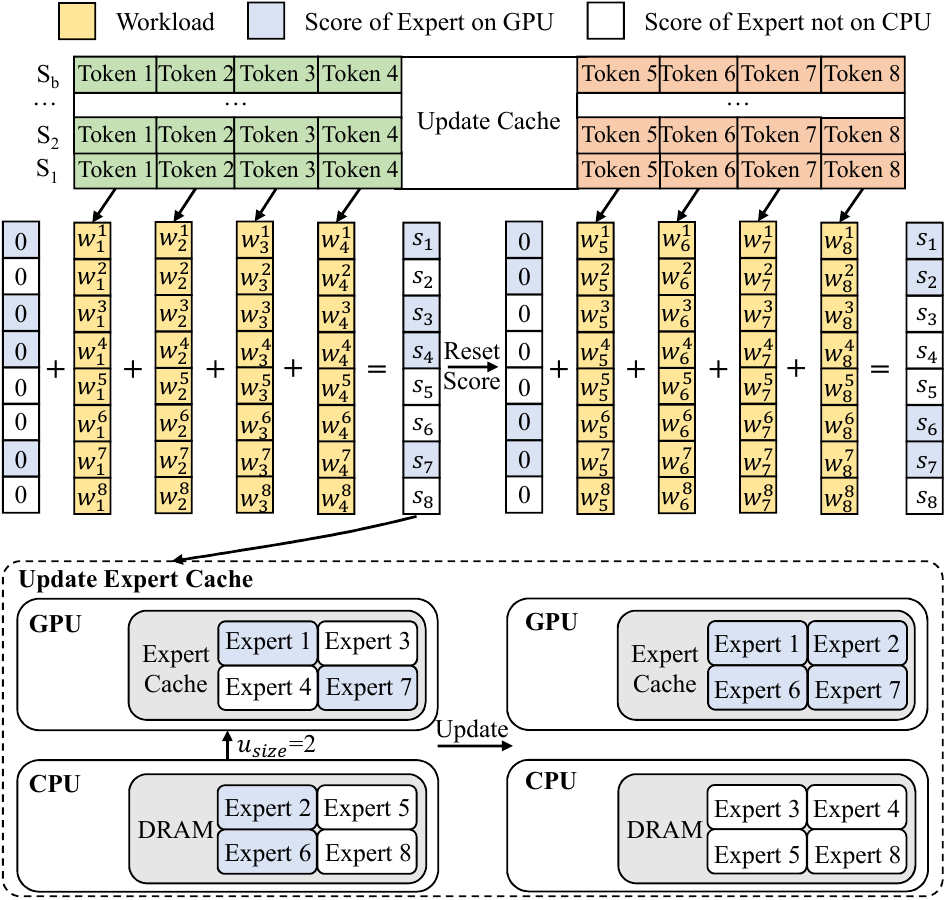}
  \vspace{-0.2cm}
  \caption{Workload-Aware Cache Replacement Strategy. Here, 
  $cache\_size=4$, $w_{\text{size}}=4$, $u_{\text{size}}=2$, 
  each layer totally has 8 experts, and $S_i$ denotes the $i$-th 
  sequence in a batch of size $b$.}
  \vspace{-0.2cm}
  \label{cache_method}
\end{figure}

Figure~\colorref{cache_method} and 
Algorithm~\colorref{alg:workload_cache} illustrate the cache workflow within a 
single MoE layer with our proposed Workload-Aware Cache Replacement strategy. 
Each MoE layer maintains $n$ experts ($cache\_size$) in GPU memory and 
performs cache replacement independently following this procedure.
All experts also reside in CPU memory.
Initially, we set the workload score of each expert to zero and 
track expert IDs using two sets: ${expert\_on\_gpu}$ for those experts currently 
cached on GPU, and ${expert\_on\_cpu}$ for those not on GPU, 
as shown in lines 1-3.
We then define a sliding token window of size $w_{\text{size}}$. 
Within each window, the strategy obtains the workloads of experts 
for each token (line 5) and updates workload scores accordingly (line 6). Specifically, 
the accumulated workload score $s_k$ for expert $k$ is calculated as:
\begin{equation}
\setlength{\abovedisplayskip}{3pt}
\setlength{\belowdisplayskip}{3pt}
s_k = \sum_{i=1}^{w_{\text{size}}} \text{workload}_i^k \ \text{,}
\label{eq:workload_accumulation}
\end{equation}
where $\text{workload}_i$ 
is an $N$-dimensional vector representing the workload distribution of $N$ experts 
when processing the $i$-th 
token in the current window and $\text{workload}_i^k$ denotes the 
workload of expert $k$.
After processing a window of $w_{\text{size}}$ tokens (line 9), 
we perform the cache replacement: 
we select the $u_{\text{size}}$ experts with the highest 
scores from the CPU (line 10) and the $u_{\text{size}}$ experts with 
the lowest scores from the GPU (line 11). 
Selected experts on GPU are replaced by the selected experts on CPU
to maximize cache 
utility.
After the replacement, we update the ${expert\_on\_gpu}$ and ${expert\_on\_cpu}$,
and the scores ($s$) are reset to zero. Then,
the strategy continues performing the above process using newly generated tokens until 
the inference process is complete.

Moreover, the expert cache can cooperate with our Greedy Assignment strategy: if an expert is
already resident on the GPU, its PCIe
transfer cost is treated as zero, and only its GPU compute time is counted during
scheduling.

\begingroup
\tighttextfloatsepOnce{10pt plus 0pt minus 0pt}
\begin{algorithm}[t]
  \caption{Workload-Aware Cache Replacement}
  \label{alg:workload_cache}
  \begin{algorithmic}[1]
  \State $s = [0, 0, \dots, 0]^N$ 
  \State ${expert\_on\_gpu} =  \{e^{\text{gpu}}_1, e^{\text{gpu}}_2, \dots, e^{\text{gpu}}_n\}$
  \State ${expert\_on\_cpu} =  \{e^{\text{cpu}}_1, e^{\text{cpu}}_2, \dots, e^{\text{cpu}}_m\}$
  \For{$i = 0$ to $\text{max\_length} - 1$}
      \State $\text{workload}_i = \text{get\_workload}(x_i)$
      \State $s = s + \text{workload}_i$
      \If{$\text{token}_i == \text{EOS}$}
          \State \textbf{break}
      \ElsIf{$i \bmod w_{\text{size}} == 0$}
          \State ${cpu\_trans\_ind} = \text{TopK}(s[{expert\_on\_cpu}])$
          \State ${gpu\_evict\_ind} = \text{TopK}(s[{expert\_on\_gpu}])$
          \State \text{Evict experts on GPU with indices in} ${gpu\_evict\_ind}$
          \State Transfer experts from CPU to GPU with indices in ${cpu\_trans\_ind}$
          \State \text{Update} ${expert\_on\_gpu}$ \text{and} ${expert\_on\_cpu}$
          \State $s = [0, 0, \dots, 0]^N$ 
      \EndIf
  \EndFor
  \end{algorithmic}
  \end{algorithm}
\endgroup

\begin{figure*}[t]
  \centering
  \vspace{-0.4cm}
  \includegraphics[scale=0.602,trim=0cm 0cm 0cm 0cm,clip]{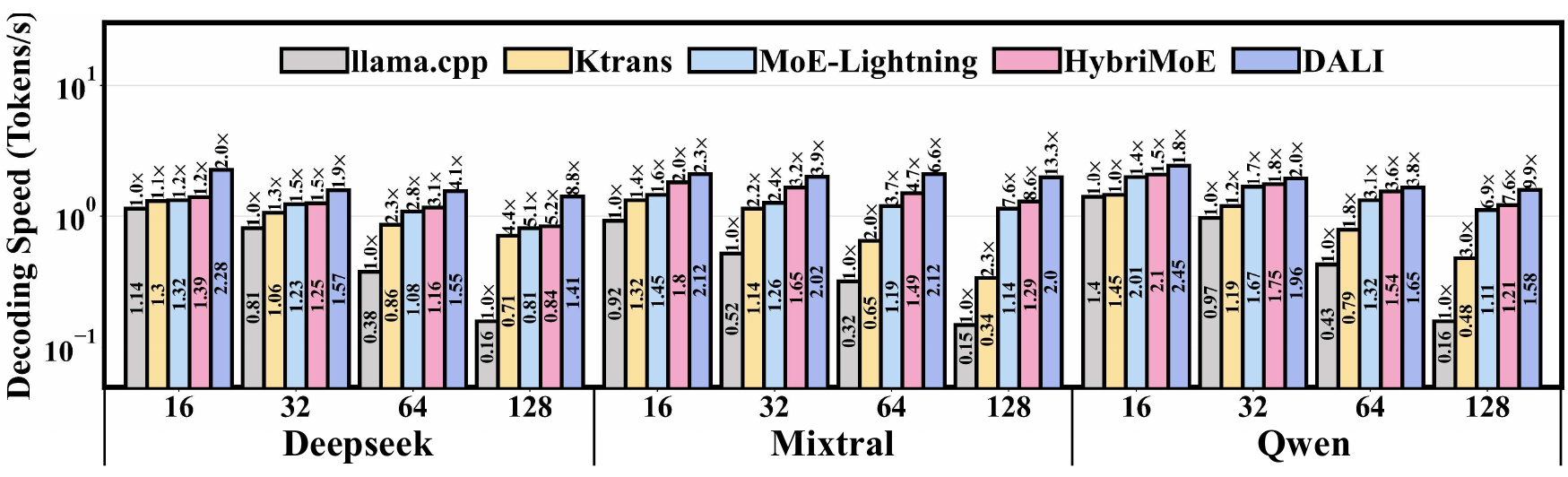}
  \vspace{-0.7cm}
  \caption{Comparison of decoding 
  speed across different models and batch sizes. For all models, the cache ratio is 50\%
  and $w_{size}=4$. For Mixtral, $u_{size}=1$ and the prefetch size is 1. For DeepSeek and Qwen,
  $u_{size}$=8 and the prefetch size is 4.}
\label{overall_results}
\end{figure*}

\vspace{-0.3cm}
\section{Implementation}
\vspace{-0.1cm}

We implement our proposed DALI framework based on the open-source KTransformers 
framework\colorcite{ktransformers}.
To support expert-wise hybrid execution, we first extend it with over 
~1,000 lines of C++ and ~2,000 lines of Python code. In addition, we develop three key 
modules: \texttt{plan\_solver}, \texttt{prefetch\_tool}, and \texttt{cache\_tool}
to support our Greedy Assignment, Residual-Based Prefetching, and Workload-Aware 
Cache Replacement techniques, respectively.
All modules are encapsulated into user-friendly APIs. 

\vspace{-0.3cm}
\section{Evaluation}

\subsection{Experimental Setup}
\label{experimental_setup}

\textbf{1) Models.} We evaluate DALI on three widely-used, 
open-sourced MoE models: DeepSeek-V2-Lite-Chat (DeepSeek), 
{Qwen3-30B-A3B (Qwen)}, and Mixtral-8x7B-Instruct (Mixtral). 
The details of the used MoEs are 
summarized in Table~\colorref{config_of_moe}. 

{\textbf{2) Datasets.}} We evaluate on two standard LLM datasets: 
{C4}\colorcite{raffel2020exploring}
and {Wikitext}\colorcite{merity2016pointer}. 
To construct the residual vector used in Residual-Based 
Prefetching, we sample 1K sequences from Wikitext to form a calibration dataset 
and perform 
inference to collect token-level features. Then we can obtain the residual vector as 
Equation~\colorref{eq:residual_matrix}.
For speed benchmarking, we sample 
input sequences from the C4 dataset.

{\textbf{3) Baselines.}} We compare with four state-of-the-art 
MoE offloading frameworks: {llama.cpp}~\colorcite{llama.cpp}, 
{KTransformers}~\colorcite{chen2025ktransformers}, MoE-Lightning~\colorcite{cao2025moe}, 
and {HybriMoE}~\colorcite{zhong2025hybrimoe}. 
Ktransformers and
llama.cpp are layer-wise hybrid frameworks and assign MoE layers to either the CPU or GPU. 
MoE-Lightning searches the optimal model deployment strategy before inference 
based on its proposed performance analysis model. 
HybriMoE is an expert-wise hybrid framework 
and incorporates 
both expert prefetching and caching techniques. 
To ensure a fair comparison, we set the number of CPU cores to 16 and threads to 32 for 
all frameworks. 
We also ensure that all frameworks use comparable GPU memory.
For HybriMoE and DALI, we cache the same number of experts 
on GPU.
For MoE-Lightning, llama.cpp and KTransformers, which do not support expert caching on GPU, 
we control the number of MoE layers stored and executed on the GPU in these frameworks to 
maintain a comparable memory usage with that of DALI and HybriMoE.
\begin{table}[t]
  \vspace{-0.1cm}
  \caption{The configuration of the used MoE architectures.}
  \normalsize
  \vspace{-0.3cm}
  \label{config_of_moe}
  \begin{center}
    \begin{tabular}{cccc}
      \toprule[1.5pt]
                           & DeepSeek       & Qwen           & Mixtral         \\ \bottomrule[1pt]
      Layers               & 27             & 48             & 32              \\
      Hidden size         & 2048           & 2048           & 4096            \\
      Shared   Experts per Layer    & 2              & 0              & 0               \\
      Routed Experts per Layer     & 64             & 128             & 8               \\
      Activated   Experts  & 6              & 8              & 2               \\\bottomrule[1.5pt]
      \end{tabular}
\end{center}
\vspace{-0.4cm}
\end{table}

{\textbf{4) Metrics.}} 
The inference of MoE models is divided into the prefill 
and decoding phases. We evaluate the two 
phases separately (\textit{prefill speed} and \textit{decoding speed}), 
using tokens per second (tokens/s) as the performance metric. 
The average speed across all sequences in a batch is reported as the speed metric of this batch.
If not specific, for prefill 
benchmarks, the prompt length is set to 64, and 
for decoding benchmarks, we set the prompt length and generated lengths both to 64.

{\textbf{5) Hardware Platform.}} 
All experiments are conducted on a platform equipped with an
AMD EPYC 7532 CPU with 64 cores, 256GB DDR4 DRAM,
an NVIDIA RTX 3090 GPU (24GB memory), and
PCIe 4.0 $\times$16 interface.

\begin{figure}[t]
  \centering
  \vspace{-0.4cm}
  \includegraphics[scale=0.32,trim=0cm 0cm 0cm 0cm,clip]{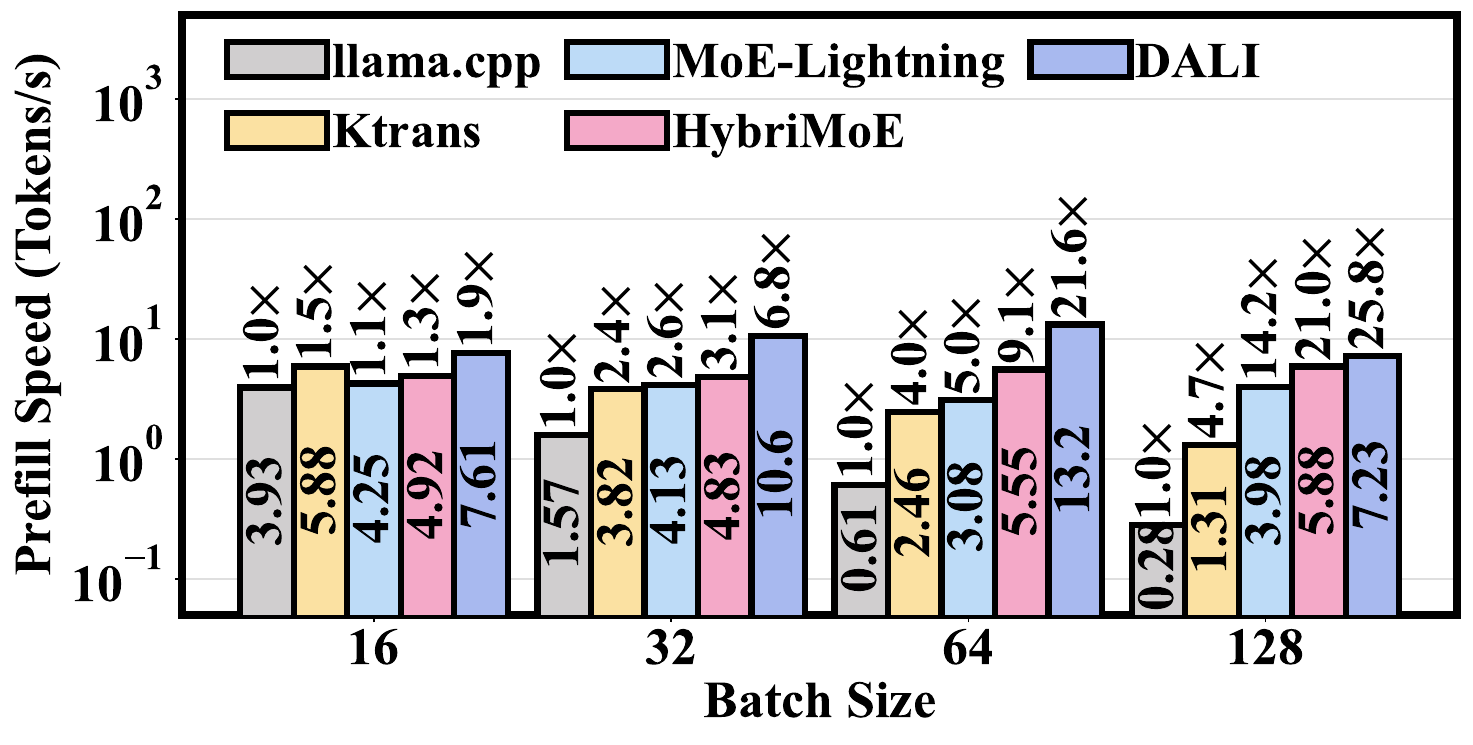}
  \vspace{-0.5cm}
  \caption{Prefill speed on DeepSeek 
  under varying batch sizes.}
  \vspace{-0.4cm}
  \label{prefill_cmp}
\end{figure}

\subsection{Overall Results}

As shown in Figure~\colorref{overall_results}, we compare decoding speed under various batch sizes. 
Since Fiddler is much slower than DALI (on average 14.3$\times$), we omit its detailed comparisons. 
Compared with {llama.cpp}, {KTransformers}, MoE-Lightning, and HybriMoE, 
DALI achieves average speedups of 3.97$\times$, 2.16$\times$, 1.48$\times$, and 1.32$\times$, respectively. 
The speedup over 
KTransformers and llama.cpp is particularly significant, mainly because our Greedy 
Assignment strategy enables effective utilization of both CPU and GPU resources for 
MoE computation. In contrast, KTransformers and llama.cpp 
uses a static layer-wise mapping that prevents parallel 
execution, failing to leverage heterogeneous hardware effectively, especially when 
processing high-workload experts. 
Compared with MoE-Lightning, DALI avoids numerous 
asynchronous transfers and frequent stream switches by combining accurate prefetching 
and caching, and MoE-Lightning's fixed CPU/GPU placement before inference makes it poorly 
suited to MoE's dynamic workload patterns. Compared with HybriMoE, DALI further improves 
performance through dynamic expert planning to better utilize CPU-GPU parallelism, 
while our {Residual-Based Prefetching} and {Workload-Aware Cache Replacement} 
strategies significantly reduce PCIe communication overhead, providing additional speedup.

Figure~\colorref{prefill_cmp} presents the prefill speed under different batch 
sizes for DeepSeek. On average, DALI achieves speedups of 7.62$\times$, 3.80$\times$, 2.45$\times$
and 2.00$\times$ over llama.cpp, KTransformers, MoE-Lightning, and HybriMoE, respectively. 
The performance gain over KTransformers and llama.cpp is substantial, largely due to their 
limited use of GPU resources. Since the CPU is not well-suited for high-workload tasks, 
especially at large batch sizes, their performance deteriorates quickly. Compared to HybriMoE, 
our framework also obtains significant performance improvements thanks to our dynamic expert assignment
strategy and the optimization on prefetching and caching.

\begin{figure}[t]
  \centering
  \vspace{-0.4cm}
  \includegraphics[scale=0.35,trim=0cm 0cm 0cm 0cm,clip]{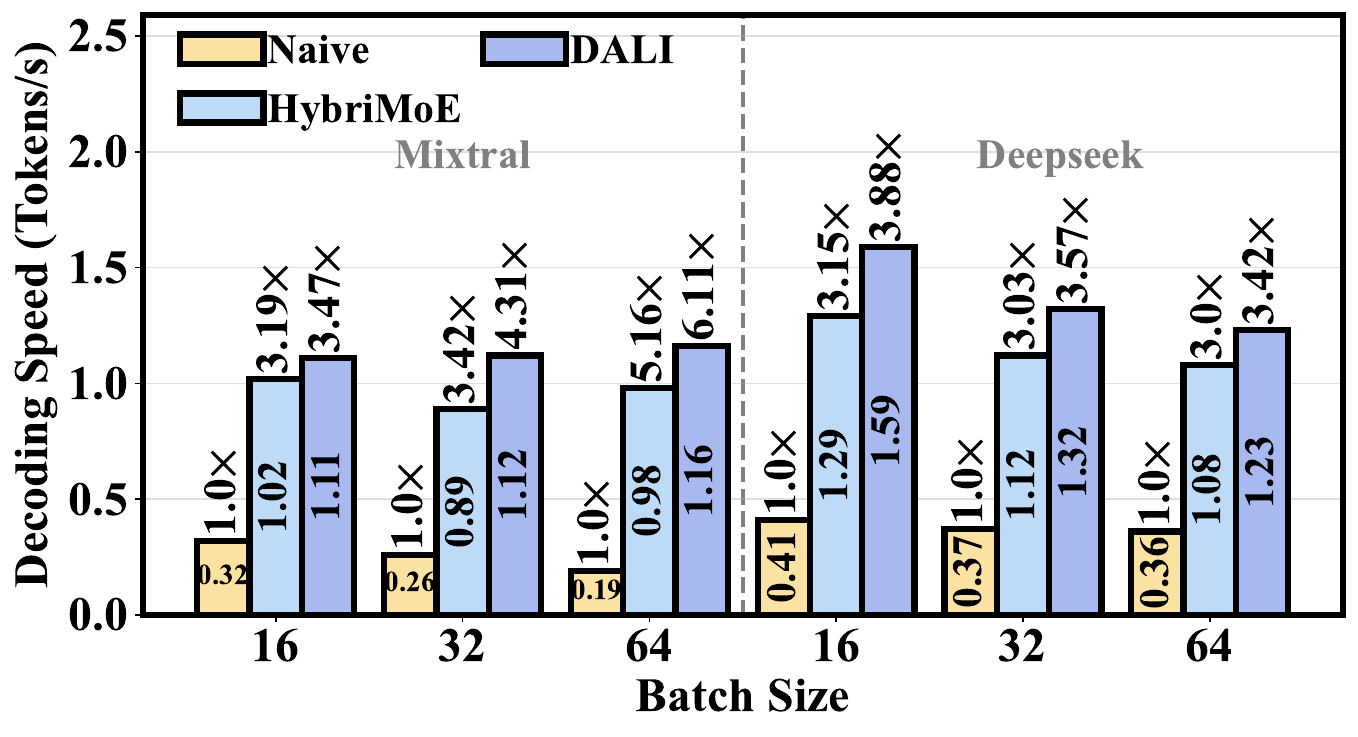}
  \vspace{-0.5cm}
  \caption{Comparison of decoding speed between HybriMoE 
  and DALI using their respective assignment strategies. 
  }
  \vspace{-0.4cm}
  \label{plan_cmp_baseline}
\end{figure}

\begin{figure}[t]
  \begin{minipage}[]{0.22\textwidth}
    \centering
    \includegraphics[scale=0.28,trim=0cm 0cm 0cm 0cm,clip]{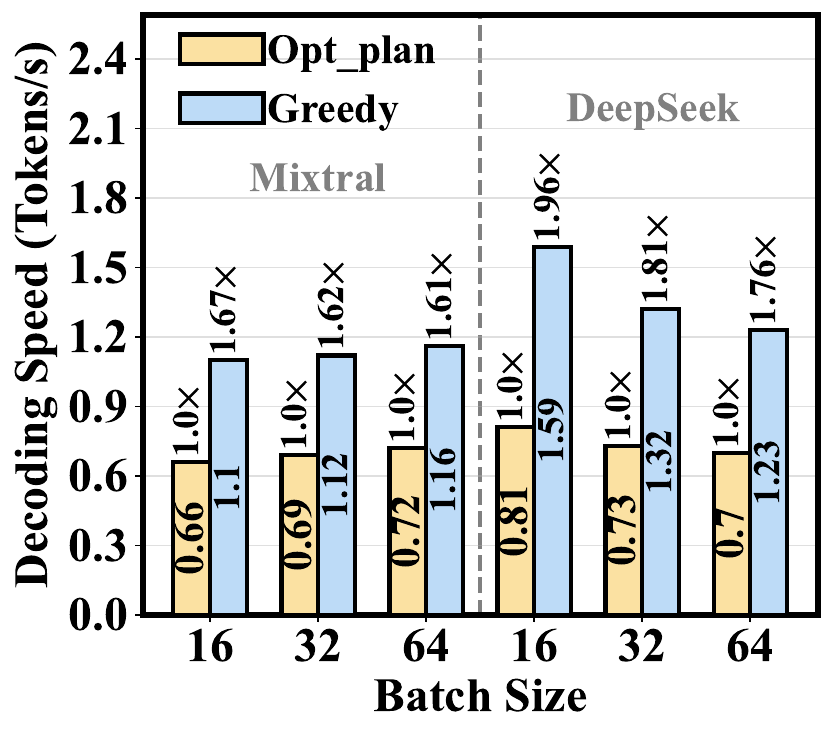}
    \vspace{-0.7cm}
    \caption{Decoding speed comparison between greedy 
    assignment and optimal assignment strategies.}
    \label{plan_opt_cmp}
      \vspace{-0.2cm}
    \end{minipage}
  \qquad
  \
  \begin{minipage}[]{0.2\textwidth}
    \vspace{-0.5cm}
    \captionof{table}{MoE spent time (s) comparison 
    between different expert assignment strategies. Decoding length = 32.}
    \vspace{-0.2cm}
    \label{plan_opt_time_cmp}
    \scriptsize
    \begin{tabular}{p{0.4cm}<{\centering}p{0.8cm}<{\centering}p{1.4cm}<{\centering}}
      \toprule[1.5pt]
      \begin{tabular}[c]{@{}c@{}}Batch \\ Size\end{tabular} & Opt\_plan & Greedy          \\ \bottomrule[1pt] 
      \multicolumn{3}{c}{DeepSeek}             \\ \hline
      16         & 12.4      & 14.4   (↓14\%)  \\
      32         & 19.7      & 22.4   (↓12\%)  \\ \hline
      \multicolumn{3}{c}{Mixtral}              \\ \hline
      16         & 22.6      & 26.7 (↓15\%)     \\
      32         & 28.0      & 30.4 (↓7.8\%) \\ \bottomrule[1.5pt] 
      \end{tabular}
      \vspace{-0.2cm}
 \end{minipage}
  \vspace{-0.3cm}
\end{figure}

\subsection{Breakdown Analysis}
\label{breakdown_sec}

To analyze where our gains come from and demonstrate 
the effectiveness of our three proposed techniques separately,
we conduct more detailed breakdown analyses.

\textbf{1) Benefit of Greedy Assignment Strategy.}
Figure~\colorref{plan_cmp_baseline} shows the speedup achieved by our Greedy Assignment 
strategy. 
To isolate the impact of assignment policies, both
HybriMoE and DALI are configured to use only their assignment strategies
(i.e., without prefetching and caching techniques).
Compared to ``Naive", 
which computes all experts on the CPU without any scheduling strategy, 
HybriMoE and our greedy strategy achieve average speedups of 3.58$\times$ 
and 4.42$\times$, respectively. 
Furthermore, our greedy strategy outperforms HybriMoE's by 23\% because our greedy assignment
achieves better load balance between the CPU and GPU and fully exploits the 
heterogeneous hardware resources.

Although solving the 0-1 optimization problem yields the optimal expert 
assignment, its runtime solving cost is prohibitively high. To balance 
efficiency and quality, we develop a heuristic Greedy Assignment. 
As shown in Figure~\colorref{plan_opt_cmp}, our greedy strategy achieves a 
1.70$\times$ speedup over the ``Opt\_plan'' method, which first solves the 
0-1 optimization problem and performs the MoE computation according to the optimal 
expert assignment. While 
``Opt\_plan'' provides a theoretically optimal assignment, its solving 
overhead largely diminishes its acceleration benefit. In contrast, the 
greedy strategy delivers near-optimal assignments with negligible cost: 
the latency overhead introduced by the greedy strategy 
is only 4.5\% (v.s.\ 55\%) of the total inference time. Moreover, as 
shown in Table~\colorref{plan_opt_time_cmp}, when comparing only the MoE 
execution time (excluding solving costs), the greedy schedule attains 
up to 92\% of the performance of the optimal solution.

\begin{figure}[t]
  \vspace{-0.6cm}
  \centering
  \subfloat[]{
      \begin{minipage}[t]{0.48\linewidth}
        \centering
        \includegraphics[scale=0.28,trim=0cm 0cm 0cm 0cm,clip]{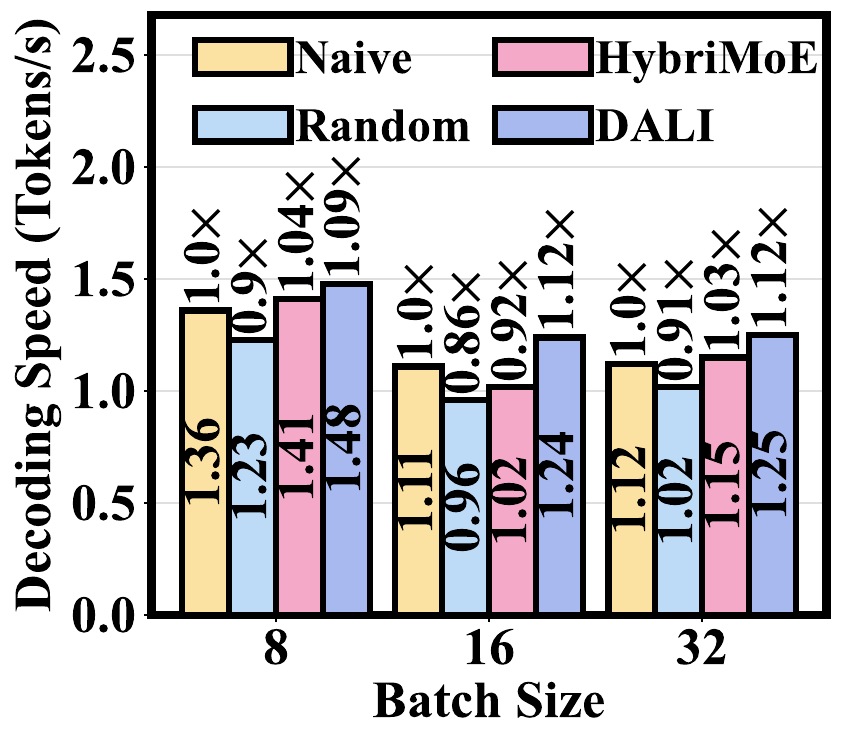}
        \vspace{-0.4cm}
        \label{prefetch_cmp}
      \end{minipage}
  }
  \hspace{0.2cm}
   \subfloat[]{
      \begin{minipage}[t]{0.42\linewidth}
        \centering
        \includegraphics[scale=0.28,trim=0cm 0cm 0cm 0cm,clip]{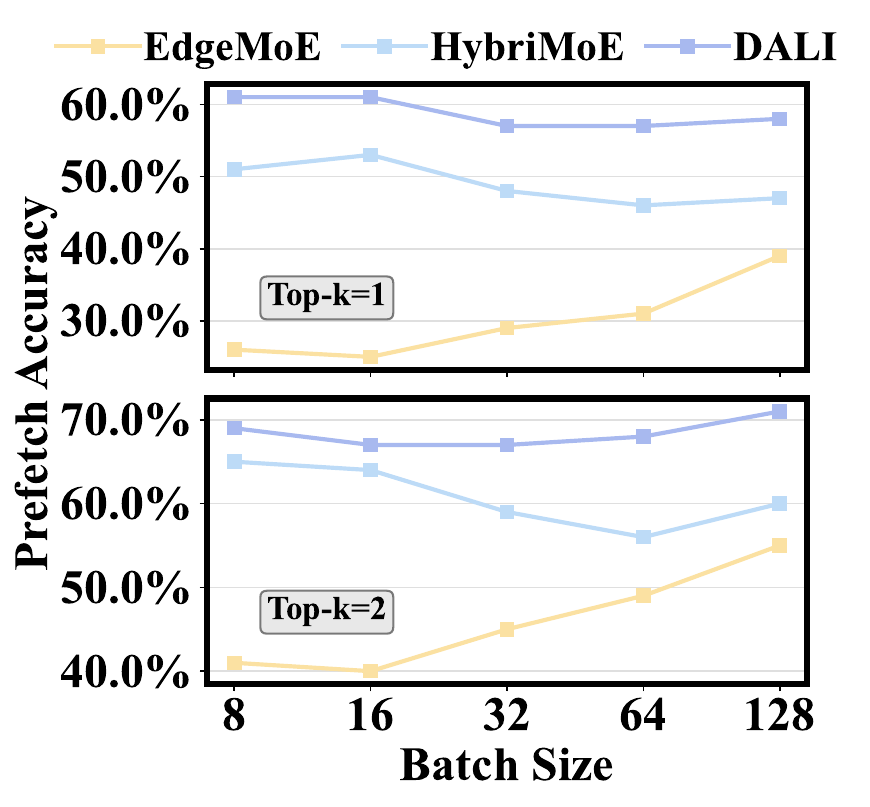}
        \vspace{-0.6cm}
        \label{prefetch_acc_mixtral}
      \end{minipage}
   }
   \vspace{-0.4cm}
   \caption{(a) Speedup comparison of different prefetching methods. 
   Each method prefetches two experts. (b) Prefetch accuracy comparison using different methods on Mixtral. 
   \textit{Top-k} indicates prefetching the top-k highest-workload experts.}
   \label{prefetch_cmp_fig}
   \vspace{-0.2cm}
\end{figure}

\begin{figure}[t]
  \vspace{-0.6cm}
  \centering
  \subfloat[]{
      \begin{minipage}[t]{0.55\linewidth}
        \centering
      \includegraphics[scale=0.32,trim=0cm 0cm 0cm 0cm,clip]{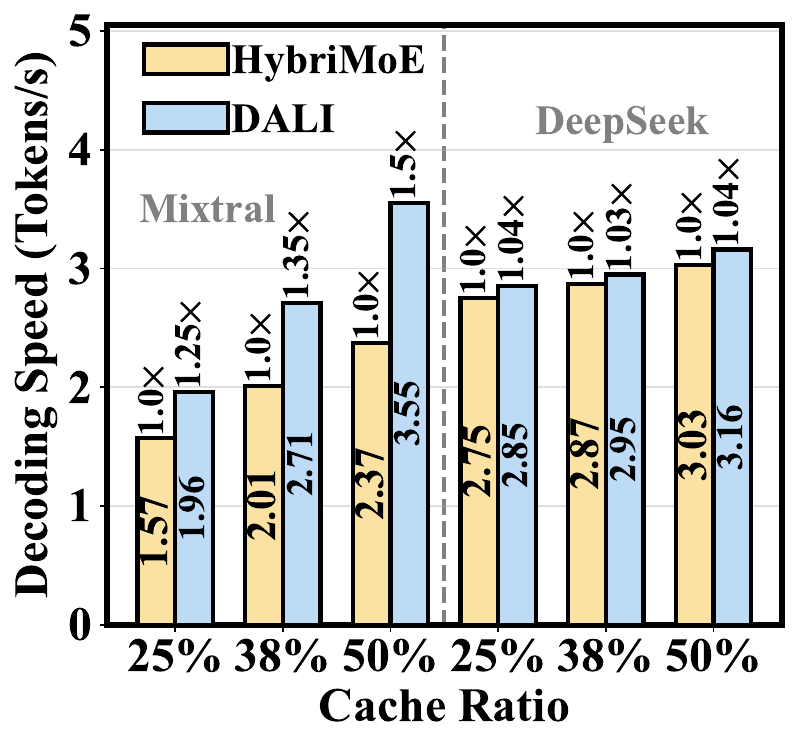}
      \vspace{-0.2cm}
      \label{cache_speedup_cmp}
      \end{minipage}
  }
   \subfloat[]{
      \begin{minipage}[t]{0.37\linewidth}
        \centering
        \includegraphics[scale=0.32,trim=0cm 0cm 0cm 0cm,clip]{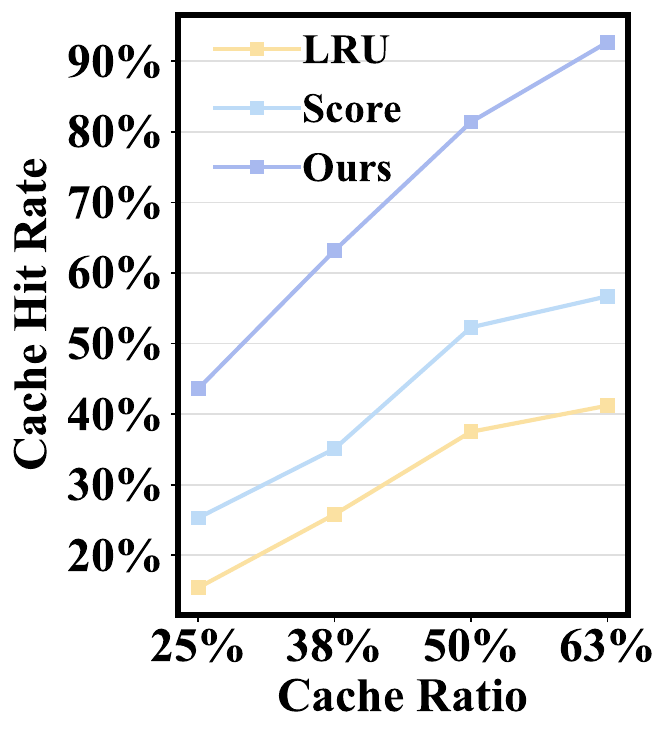}
        \vspace{-0.4cm}
        \label{hit_rate_cmp_mixtral}
      \end{minipage}
   }
   \vspace{-0.4cm}
   \caption{(a) Decoding speed comparison between HybriMoE's update strategy and our 
   workload-aware strategy under different cache ratios. Batch size is 4. 
   (b) Cache hit rate comparison across different update strategies on 
   Mixtral. Batch size is 4.}
   \label{cache_cmp}
   \vspace{-0.5cm}
\end{figure}

To further illustrate the effect of our Greedy Assignment on balancing workload, 
we compare CPU and GPU
execution times of HybriMoE and DALI in Appendix~\colorref{workload_analyses_appendix}. Moreover, we provide 
a more detailed analyses on the latency overhead introduced by our greedy strategy in 
Appendix~\colorref{overhead_sec}. DALI develops
a Greedy Assignment strategy to obtain a near\mbox{-}optimal assignment while significantly reducing
the solving cost. In addition, we also explore the effects of other approximate
solving methods, such as beam search, in Appendix~\colorref{other_appendix}.

\begin{figure*}
  \centering
  \vspace{-0.5cm}
  \subfloat[]{\includegraphics[scale=0.31]{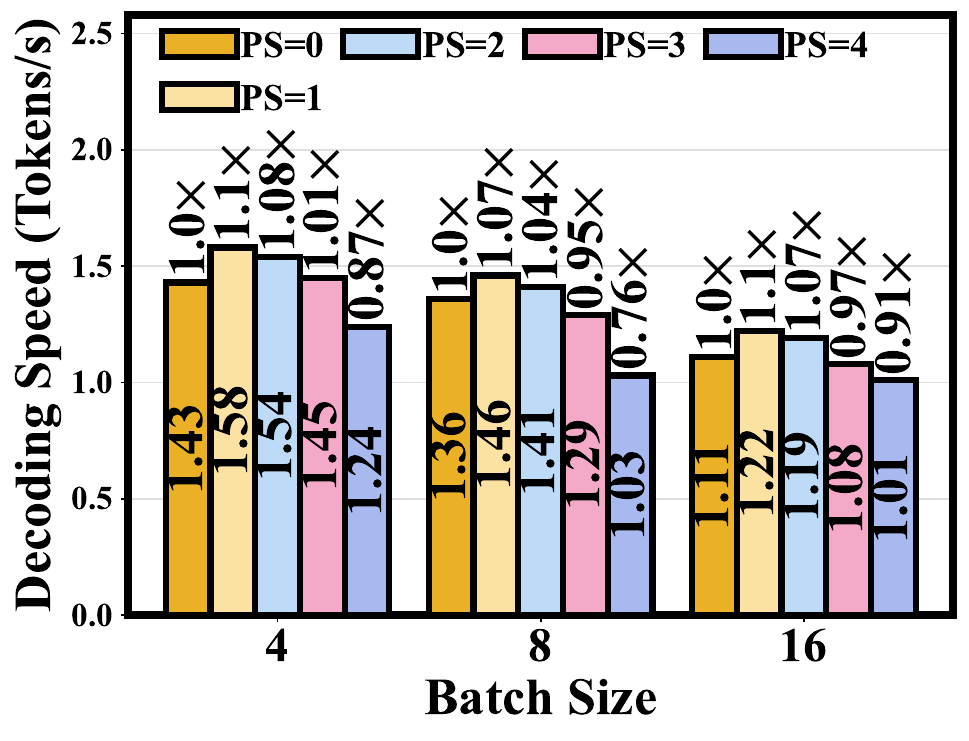}\label{prefetch_size_various}} 
  \subfloat[]{\includegraphics[scale=0.31]{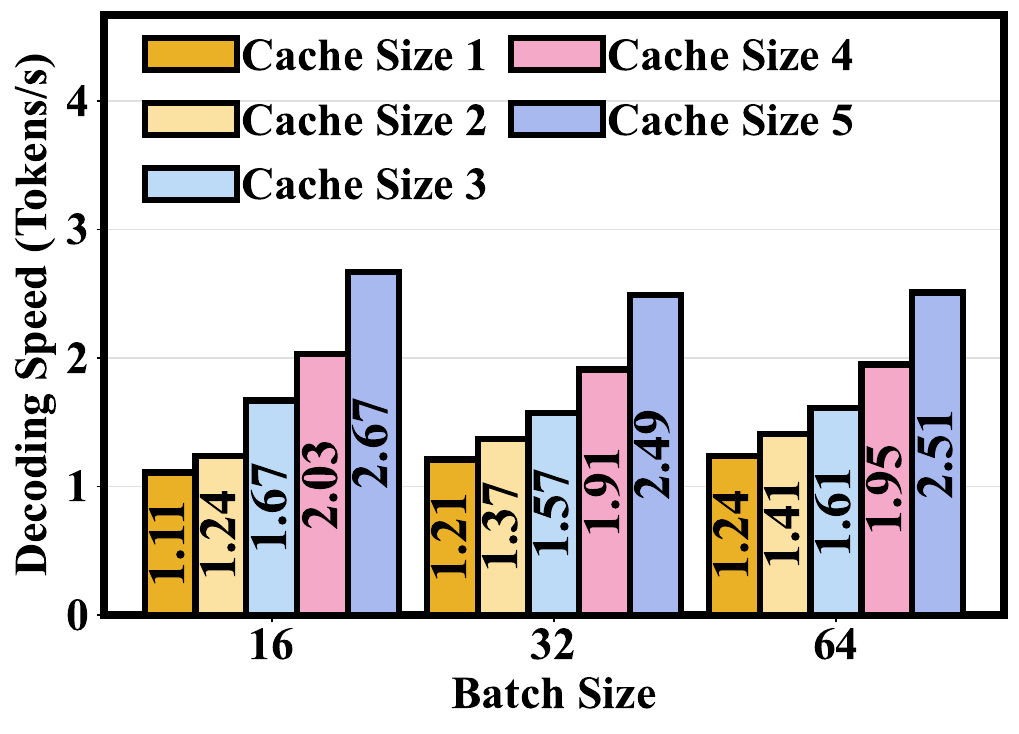}\label{cache_size_speedup_3}} 
  \subfloat[]{\raisebox{0.1cm}{\includegraphics[scale=0.31]{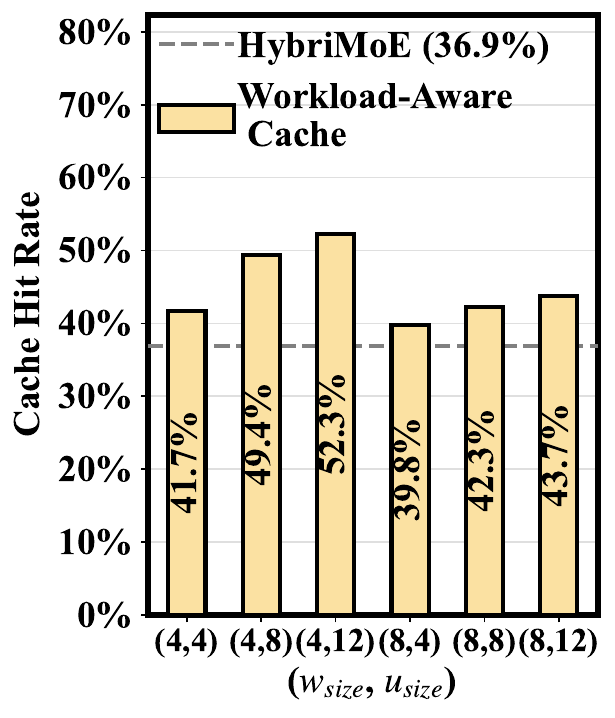}}\label{wsize_usize_cmp}}
  \subfloat[]{\raisebox{0.05cm}{\includegraphics[scale=0.31]{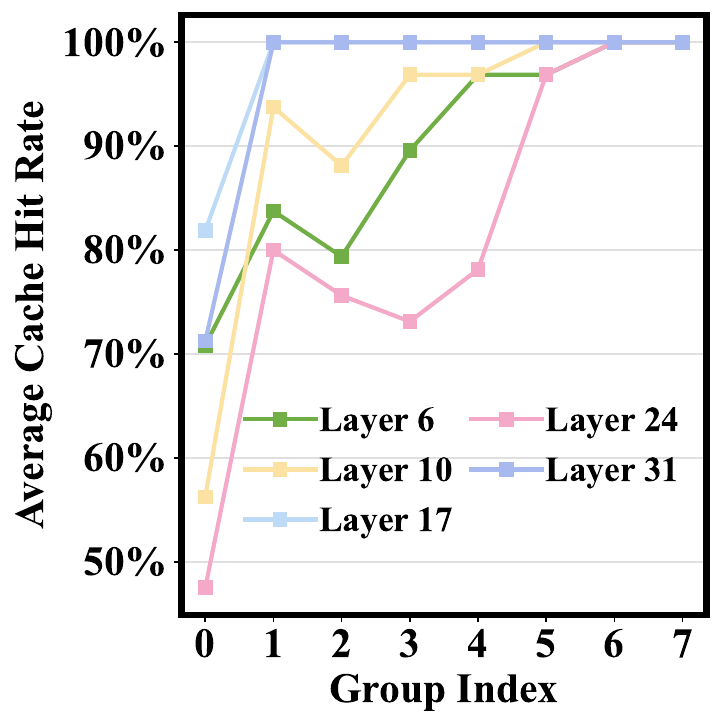}}\label{hit_rate_various_4}}
  \vspace{-0.4cm}
  \caption{{(a) Decoding speed on Mixtral with varying numbers of prefetched experts (`PS' denotes the prefetch size).} 
  (b) Decoding speed on Mixtral with varying numbers of cached experts on GPU.
  (c) Cache hit rate of the Workload-Aware Cache Replacement strategy on DeepSeek under 
  different $w_{\text{size}}$ and $u_{\text{size}}$ configurations. Batch size is 4.
  (d) Cache hit rate changes across layers during token generation on Mixtral, when caching 
  4 experts per layer. Batch size = 4, decoding length = 64, $w_{\text{size}} = 8$, 
  $u_{\text{size}} = 1$. 
  We report the average hit rate of every 8 tokens as a group.} 
  \label{scale_results} 
  \vspace{-0.4cm}
\end{figure*} 

\begingroup
\begin{figure}[t]
  \centering
  \includegraphics[scale=0.36,trim=0cm 0cm 0cm 0cm,clip]{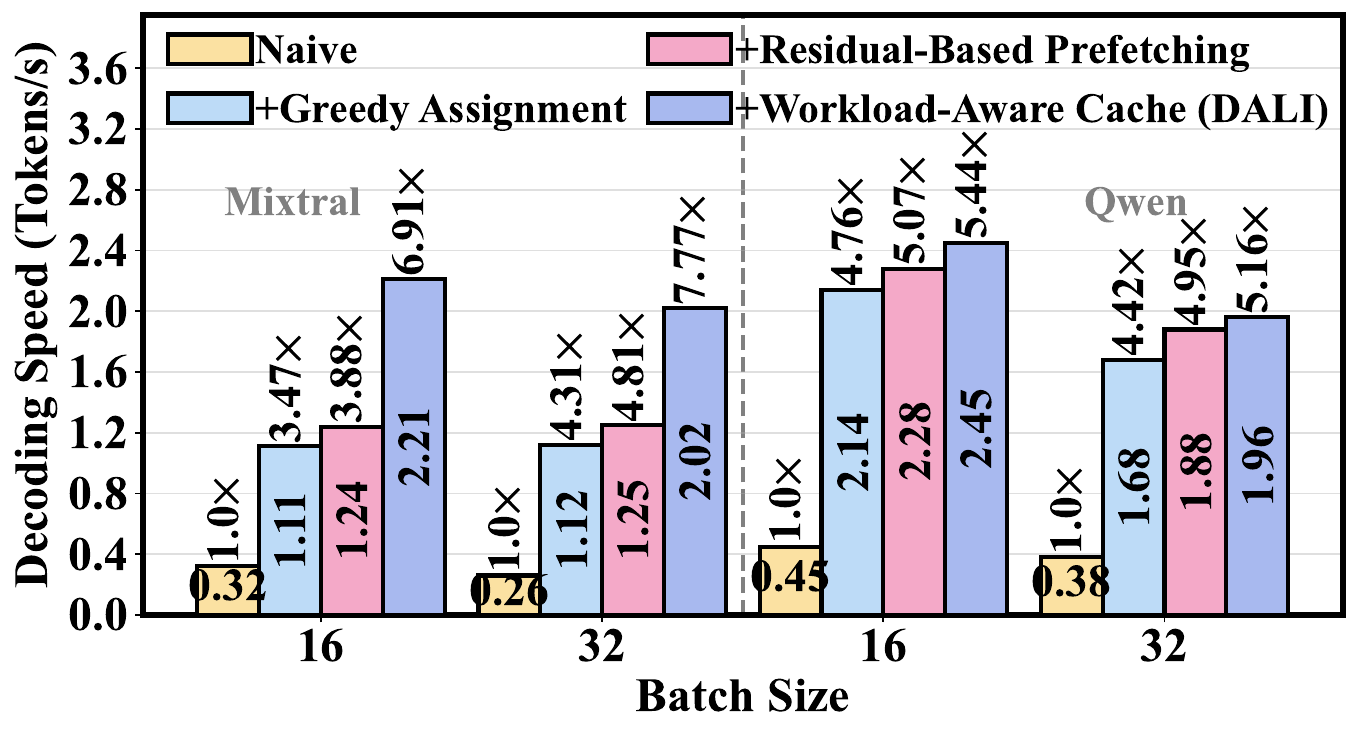}
  \vspace{-0.5cm}
  \caption{Breakdown analysis of performance gains. 
  In each group, from the second bar, 
  the right bar includes one more optimization than the adjacent left bar. 
  {``Naive" refers to the case without any optimization and
  offload all experts on CPU using Ktransformers.}
  Cache ratio is 25\%, prefetch size is 1 (Mixtral) and 8 (Qwen).}
  \vspace{-0.6cm}
  \label{breakdown_exp}
\end{figure}
\endgroup

\textbf{2) Benefit of Prefetching.}
Figure\colorref{prefetch_cmp} evaluates various prefetching strategies on
Mixtral. ``Naive'' denotes the baseline using only Greedy Assignment
(built on KTransformers), ``Random'' performs random prefetching, and
``HybriMoE'' adopts its own prefetching method. ``Random'' performs
worse than ``Naive'' due to frequent stalls caused by incorrect
prefetches. HybriMoE provides modest improvement but suffers from low
accuracy in identifying high-workload experts. In contrast, our
Residual-Based Prefetching significantly improves prediction accuracy,
resulting in larger speedups.
Figure\colorref{prefetch_acc_mixtral} further compares prefetch accuracy 
on
Mixtral. Our method consistently achieves the highest accuracy 
on high-workload experts.
To explain why our Residual-Based Prefetching is effective, we 
provide a further analyses from the perspective of the cosine
similarity in 
Appendix~\colorref{prefetch_analysis_appendix}.

\textbf{3) Benefit of Cache Replacement Strategy.}
Figure~\colorref{cache_speedup_cmp} compares the inference speed when employing different 
GPU cache replacement strategies. The cache ratio denotes the proportion of experts 
cached on the GPU. Compared to HybriMoE, which replaces cache based on expert 
activation score, our Workload-Aware Cache Strategy achieves a 1.23$\times$ speedup. 
This is because our strategy more effectively captures the expert utilization in 
dynamic workload scenarios, 
resulting in higher cache hit rates and reduced PCIe traffic.
Figure~\colorref{hit_rate_cmp_mixtral} shows the comparisons on cache hit 
rates under different cache ratios 
when applying various replacement strategies. Our workload-aware method consistently 
outperforms both LRU and HybriMoE's score-based approach.

\textbf{4) Overall Breakdown.}
Figure~\colorref{breakdown_exp} presents the individual performance 
gains from each technique. Compared to ``Naive", the Greedy Assignment 
Strategy delivers a 4.1$\times$ speedup—the most significant among the three. 
This is because it maximizes heterogeneous resource utilization by intelligently 
assigning experts across CPU and GPU.
Prefetching yields a marginal 9\% gain, mainly due to two reasons: 
(1) prefetching requires additional gating computations for prediction, and (2) 
it incurs CUDA stream switching overhead, which both partially diminish the benefits.
Finally, our cache technique contributes a further 38\% speedup, as our 
Workload-Aware Cache Replacement strategy improves cache hit rate, 
thereby reducing PCIe communication overhead and accelerating overall inference.

\subsection{Sensitivity Analysis}
\label{sens_sec}

In this section, we explore how the inference speed of our DALI varies 
with key parameters such as 
prefetch size, 
cache ratio, $w_{\text{size}}$, and $u_{\text{size}}$.

\textbf{1) Effect of Prefetch Size on Inference Speed.}
Figure~\colorref{prefetch_size_various} shows the inference speed on Mixtral 
as we vary the number of experts being prefetched. We observe that prefetching 
only one expert—the one with the highest predicted workload—yields the best performance. 
The reasons are twofold: (1) the expert with the highest workload predicted by residual-based 
prefetching is often actually used on the GPU, thus reducing the prediction error 
rate, and (2) as
more experts are prefetched, the computation time becomes insufficient to overlap the 
communication cost, resulting in reduced speed.

\textbf{2) Effect of Cached Expert Count on Inference Speed.}
Figure~\colorref{cache_size_speedup_3} shows decoding speed as we increase the number of 
experts cached per layer on Mixtral. The decoding speed improves 
with the increased cache size, demonstrating the scalability of our Workload-Aware Cache Replacement 
strategy with respect to cache capacity.

\textbf{3) Impact of $w_{\text{size}}$ and $u_{\text{size}}$ on Cache Hit Rate.}
Figure~\colorref{wsize_usize_cmp} presents cache hit rates under different configurations 
of $w_{\text{size}}$ and $u_{\text{size}}$. We observe that smaller $w_{\text{size}}$ 
values lead to higher hit rates, indicating that more frequent cache replacement helps 
improve cache utility. Similarly, larger $u_{\text{size}}$ values (i.e., more experts 
replaced per update) also improve the hit rate. However, frequent or large-scale 
replacement incurs substantial latency overhead, which may diminish the performance 
gains from higher cache efficiency. 
Thus, a trade-off must be maintained when tuning these 
parameters. 
Therefore, we explore how $w_{\text{size}}$ and $u_{\text{size}}$ affect inference speed, as shown in 
Appendix~\colorref{explore_appendix}.
Finally, to maximize the inference speed, we select \textbf{(4,\,8)} for Qwen and DeepSeek, and
\textbf{(4,\,1)} 
for Mixtral.

\textbf{4) Cache Hit Rate Varies as Token Generation.}
Our Workload-Aware Cache Replacement strategy updates experts based on their historical workload in 
processing previous tokens. In Figure~\colorref{hit_rate_various_4}, we analyze how 
the cache hit rate varies as a sequence is progressively generated. We find that the 
hit rate consistently increases and eventually reaches up to 100\%. This indicates that 
our strategy exhibits strong domain adaptability, progressively updating the cached experts 
to better match the current sequence, thereby improving reuse and reducing PCIe communication.

\subsection{Discussion}
\label{dis_sec}

\textbf{1) Performance under Varying Decoding Lengths.} 
To evaluate the generality of DALI across different sequence lengths, 
we set the batch size to 16 and the prompt length to 32, and measure 
decoding performance on Mixtral with decoding lengths of 128, 256, 512, and 1024.
Experimental results show that DALI achieves average speedups of 2.78$\times$, 1.96$\times$, 
and 1.47$\times$ over llama.cpp, KTransformers, and HybriMoE, respectively. 
This demonstrates that DALI consistently outperforms existing MoE offloading systems 
across a range of decoding scenarios. The more detailed experimental results are shown in 
Appendix~\colorref{various_decode_len_appendix}.

\textbf{2) Applicability to Multi-GPU Platform.}
While DALI targets MoE inference on personal computers equipped with a single CPU and GPU, 
we further evaluate its generalizability in a multi-GPU setup 
(1 CPU + 2 GPUs, as described in Section~\colorref{experimental_setup}). 
During the decoding phase, DALI achieves average speedups of 3.43$\times$, 1.87$\times$, 
and 1.32$\times$ over {llama.cpp}, {KTransformers}, and {HybriMoE}, respectively, 
demonstrating strong scalability. 
As DALI is designed for single-CPU/GPU edge environments, 
further exploration of distributed environments and high CPU-GPU bandwidth server systems (e.g., GH200) 
will be pursued as future work.

\section{Conclusion}

In this work, we propose DALI, an MoE offloading 
inference framework tailored to heterogeneous hardware and the 
characteristics of MoE models. First, we introduce a Greedy 
Assignment strategy that dynamically allocates experts across CPU and GPU to fully 
leverage their respective hardware resources. Second, to reduce PCIe transfer overhead, 
we propose a Residual-Based Prefetching method that improves prefetch accuracy by 
correcting features with cross-layer residuals. Third, we develop a Workload-Aware 
Cache Replacement strategy that updates the cached expert based on the workload history, 
obtaining significantly higher cache hit rates.
Extensive experiments compared with the prior
arts demonstrate the superiority of our DALI.
\endgroup

\appendix
\section{Appendix}

\begingroup
\begin{figure}[h]
  \centering
  \includegraphics[scale=0.4,trim=0cm 0cm 0cm 0cm,clip]{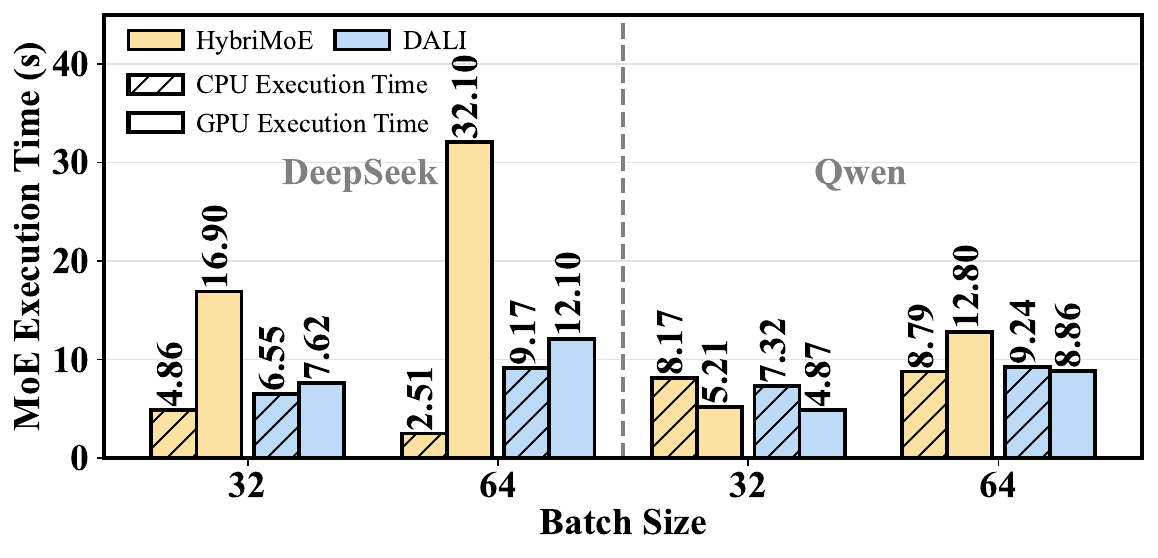}
  \vspace{-0.2cm}
  \caption{{Comparison of MoE execution time on CPU/GPU in HybriMoE 
  and DALI.}
    }
  \vspace{-0.4cm}
  \label{workload_imb_exp}
\end{figure}
\endgroup

\begingroup
\begin{table}[h]
  \small
  \caption{{The prefetch accuracy on different downstream tasks.}}
  \vspace{-0.2cm}
  \label{prefetch_acc_downstream}
  \begin{center}
    \setlength{\tabcolsep}{2.5pt}
    \begin{tabular}{ccccccc}
      \toprule[1.5pt]
                                & Method   & Arc-e           & Arc-c           & OBQA      & RTE             & Average         \\ \midrule[1pt]
      \multirow{2}{*}{DeepSeek} & HybriMoE & 75.7\%          & 75.5\%          & 69.8\%          & 69.6\%          & 72.6\%          \\
                                & DALI     & \textbf{81.1\%} & \textbf{79.4\%} & \textbf{77.4\%} & \textbf{80.1\%} & \textbf{79.5\%} \\ \hline
      \multirow{2}{*}{Qwen}     & HybriMoE & 63.2\%          & 79.0\%          & 84.7\%          & 83.6\%          & 77.6\%          \\
                                & DALI     & \textbf{93.0\%} & \textbf{94.0\%} & \textbf{92.2\%} & \textbf{93.9\%} & \textbf{93.3\%} \\ \toprule[1.5pt]
      \end{tabular}
\end{center}
\vspace{-0.4cm}
\end{table}
\endgroup

\subsection{Further Analysis on Greedy Assignment}
\label{workload_analyses_appendix}

To illustrate the effect of our Greedy Assignment, we compare CPU and GPU
execution times of HybriMoE and DALI in Figure~\colorref{workload_imb_exp}.
We can observe that DALI achieves a better load balance between 
CPU and GPU with the Greedy Assignment,
and, moreover, lowers the inference latency of the MoE layer.
Further analysis of the experimental results shows that, after applying the Greedy 
Assignment strategy, the increase in the CPU-side MoE execution time is often smaller 
than the reduction in the GPU-side MoE execution time. For example, on the DeepSeek 
model with batch size = 64, enabling Greedy Assignment increases the CPU execution 
time by 6.66\,s, while reducing the GPU execution time by 20\,s. This indicates that, 
through dynamic assignment, \textsc{DALI} offloads a portion of experts that would 
otherwise be executed on the GPU to the CPU, substantially reducing PCIe transfers 
and consequently decreasing the GPU-side MoE latency.

\begingroup
\begin{figure}[h]
  \centering
  \includegraphics[scale=0.4,trim=0cm 0cm 0cm 0cm,clip]{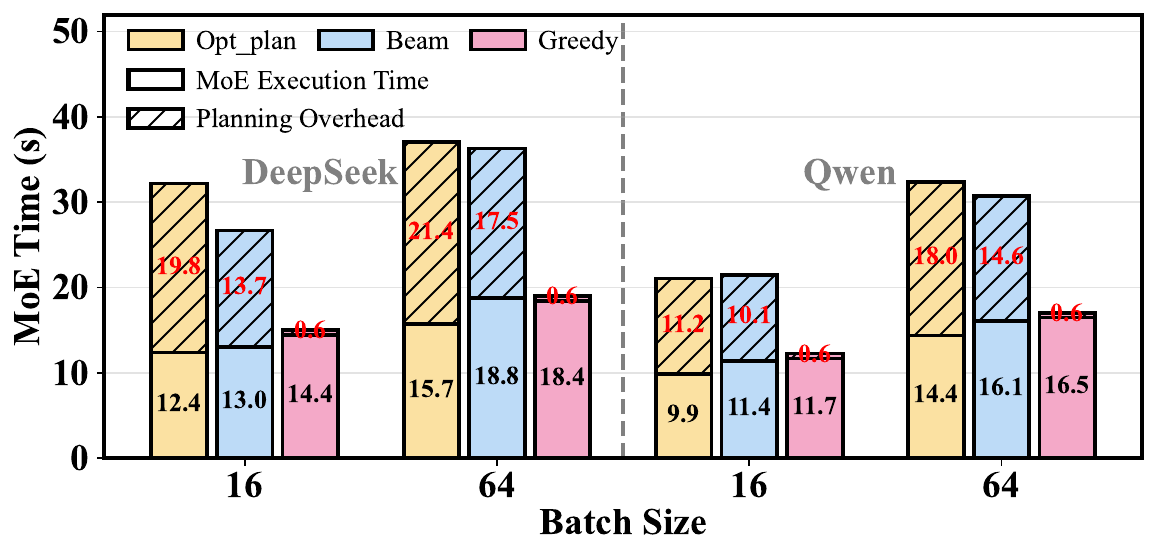}
  \vspace{-0.2cm}
  \caption{{The MoE execution time and plan overhead comparison. 
  The black number in each bar denotes the MoE execution time (without planning time) and
  the red numer denotes the planning overhead.}}
  \vspace{-0.1cm}
  \label{beam_search_cmp}
\end{figure}
\endgroup

\subsection{Try Other Sechduling Algorithms}
\label{other_appendix}

DALI develops
a Greedy Assignment strategy to obtain a near\mbox{-}optimal assignment while significantly reducing
the solving cost. In addition, we also explore the effects of other approximate
solving methods, such as beam search. 
We set the
beam size to 2 and score each beam by its execution time. In
Figure~\colorref{beam_search_cmp}, we compare the optimal plan, Greedy Assignment, and
beam search in terms of MoE execution time and planning overhead. We observe
that, although beam search can be slightly better than Greedy Assignment in some
cases, it introduces substantial solving overhead (multi\mbox{-}beam
evaluation, frequent top\mbox{-}k operations), which makes end\mbox{-}to\mbox{-}end inference
significantly slower than DALI with Greedy Assignment.

\subsection{The Generality of Residual-Based Prefetching}
\label{prefetch_general_appendix}

Our Residual-Based Prefetching strategy pre-computes the residual vector offline
on the Wikitext calibration set and reuses it for unseen tasks without fine\mbox{-}tuning.
Since Wikitext covers various types of corpora, this residual
vector remains effective on downstream tasks 
without fine\mbox{-}tuning. 
Using the residual vector derived from the
Wikitext calibration set, we compare prefetch accuracy with HybriMoE on various
downstream tasks from the EleutherAI Language Model 
Evaluation Harness~\colorcite{gao2021framework}. Compared with HybriMoE, DALI significantly improves
prefetch accuracy by 6.9\% and 15.7\% on DeepSeek and Qwen on average,
respectively, demonstrating the generalization of our Residual-Based Prefetching strategy.

\begingroup
\begin{table}[t]
  \small
  \vspace{-0.1cm}
  \caption{{The comparisons of scheduling overhead relative to end-to-end inference 
  between HybriMoE and DALI across various
  sequence lengths on DeepSeek model. Batch size is 8.}}
  \setlength{\tabcolsep}{1.5pt}
  \vspace{-0.2cm}
  \label{ga_overhead}
  \begin{center}
    \begin{tabular}{ccccccc}
      \toprule[1.5pt]
      Sequence Length         & 32     & 64     & 256    & 1024   & 2048   & Average \\ \midrule[1pt]
      HybriMoE & 2.87\% & 2.94\% & 3.02\% & 3.12\% & 3.08\% & 3.01\%  \\
      DALI     & 4.84\% & 4.54\% & 4.51\% & 4.24\% & 4.36\% & 4.50\%  \\ \bottomrule[1.5pt]
      \end{tabular}
\end{center}
\vspace{-0.2cm}
\end{table}
\endgroup

\begingroup
\begin{table}[t]
  \small
  \caption{{The memory usage comparisons 
  between HybriMoE and DALI across various batch sizes. Sequence length is 64.
 }}
 \setlength{\tabcolsep}{1.5pt}
  \vspace{-0.2cm}
  \label{mem_overhead}
  \setlength{\tabcolsep}{2.5pt}
  \begin{center}
    \begin{tabular}{ccccccc}
      \toprule[1.5pt]
                               & Method   & 8    & 16   & 32   & 64   & 128  \\ \midrule[1pt]
      \multirow{2}{*}{Mixtral} & HybriMoE & 13.4GB & 13.7GB & 14.1GB & 15.3GB & 17.8GB \\
                               & DALI     & 12.6GB & 12.8GB & 13.0GB & 13.6GB & 15.1GB \\ \hline
      \multirow{2}{*}{Qwen}    & HybriMoE & 4.79GB & 5.02GB & 5.35GB & 6.16GB & 7.42GB \\
                               & DALI     & 4.79GB & 4.98GB & 5.28GB & 5.85GB & 7.02GB \\ \bottomrule[1.5pt]
      \end{tabular}
\end{center}
\vspace{-0.2cm}
\end{table}
\endgroup

\subsection{Overhead Analysis}
\label{overhead_sec}

{In this section, we analyze DALI's potential overheads.}

{\textbf{1) Greedy Assignment Overhead.}
DALI performs Greedy Assignment to decide expert placement dynamically at runtime.
As shown in Table~\colorref{ga_overhead}, we report the scheduling overhead
relative to end-to-end
inference latency across different sequence lengths and compare it with HybriMoE's
static assignment overhead. On average, HybriMoE incurs 3.01\% overhead,
whereas our dynamic strategy incurs 4.50\%. However, as discussed in
Section~\colorref{breakdown_sec}-1, our Greedy Assignment delivers a 
4.42$\times$ end\mbox{-}to\mbox{-}end
speedup, making the 4.50\% overhead well justified.
Moreover, we observe that, because generating each token triggers a fixed number
of scheduling decisions (equal to the number of MoE layers), the fraction of
latency attributable to scheduling remains essentially constant.
}

{\textbf{2) Memory Overhead.}
Table~\colorref{mem_overhead} compares GPU memory usage between DALI and HybriMoE. We can observe that 
DALI introduces
no additional memory overhead; in fact, due to timely disposal of unused
tensors in our implementation, DALI uses less GPU memory than HybriMoE.}

\begingroup
\begin{table}[t]
  \small
  \caption{{In different layers, the comparisons on cosine similarity between the inputs used to
  predict expert activations and the ground\mbox{-}truth inputs for HybriMoE and DALI. 
  Batch size is 8.}}
  \vspace{-0.2cm}
  \label{cos_sim_cmp}
  \setlength{\tabcolsep}{1.5pt}
  \begin{center}
    \begin{tabular}{cccccccccc}
      \toprule[1.5pt]
                               & Layer ID &1 & 4    & 8    & 12   & 16   & 20                                                &23                            & Average\\ \midrule[1pt]
      \multirow{2}{*}{Qwen}    & HybriMoE &0.44 & 0.86 & 0.89 & 0.79 & 0.87 & 0.85                                              &0.83                       & 0.79\\
                               & DALI     &\textbf{0.77} & \textbf{0.95} & \textbf{0.99} & \textbf{0.93} & \textbf{0.96} & \textbf{0.97} & \textbf{0.94}    & \textbf{0.93}    \\ \hline
      \multirow{2}{*}{Mixtral} & HybriMoE &0.47 & 0.81 & 0.81 & 0.84 & 0.84 & 0.87                                              &0.88                       & 0.79\\
                               & DALI     &\textbf{0.76} & \textbf{0.86} & \textbf{0.88} & \textbf{0.92} & \textbf{0.91} & \textbf{0.92} &\textbf{0.96}     & \textbf{0.89}    \\ \bottomrule[1.5pt]
      \end{tabular}
\end{center}
\vspace{-0.2cm}
\end{table}
\endgroup

\subsection{Residual-Based Prefetching Analysis}
\label{prefetch_analysis_appendix}
{Moreover, to explain why our Residual-Based Prefetching is effective, 
we 
analyze the cosine similarity between the 
ground-truth input and the input used to predict in HybriMoE and DALI, as presented in
Table~\colorref{cos_sim_cmp}. Compared to HybriMoE, our residual\mbox{-}corrected inputs exhibit higher cosine
similarity to the ground\mbox{-}truth inputs, indicating that the correction brings
the inputs used to predict closer to the true ones. This yields higher prediction
accuracy and improves prefetch correctness.}

\begingroup
\begin{table}[t]
  \setlength{\tabcolsep}{1.5pt}
  \caption{{The inference speed (Tokens/s) under different ($w_{\text{size}}$, $u_{\text{size}}$) settings.
    Sequence length is 64 and batch size is 32.}}
  \vspace{-0.2cm}
  \label{w_u_perf}
  \begin{center}
    \begin{tabular}{ccccccc}
      \toprule[1.5pt]
            &HybriMoE   & (2, 8) & (2, 16) & (4, 8) & (4, 16) & (8, 8) \\ \midrule[1pt]
      DeepSeek & 1.25 & 1.84  & 1.76   & \textbf{1.89}  & 1.83   & 1.97  \\
      Qwen     & 1.75 & 1.88  & 1.82   & \textbf{1.96}  & 1.92   & 1.94  \\ \midrule[1pt]
            &HybriMoE   & (2,1) & (2,2)  & (4,1) & (4,2)  & (8,1) \\ \midrule[1pt]
      Mixtral  & 1.65 & 1.92  & 1.89   & \textbf{2.02}  & 1.87   & 1.98  \\ \bottomrule[1.5pt]
      \end{tabular}
\end{center}
\vspace{-0.6cm}
\end{table}
\endgroup

\subsection{Explore the Setting of $w_{\text{size}}$ and $u_{\text{size}}$}
\label{explore_appendix}

In Table~\colorref{w_u_perf}, we explore 
how $w_{\text{size}}$ and $u_{\text{size}}$ affect inference speed.
Compared with (2,\,8), the (2,\,16) setting increases $u_{\text{size}}$, i.e., the number of experts
updated per cache refresh; although it raises the cache hit rate by 4.2\%,
the extra PCIe cost outweighs the benefit, slowing the performance.
In contrast, (4,\,8) versus (8,\,8) updates the cache more frequently, but its
hit-rate gain (7.1\%) dominates, improving overall speed.
Notably, even with the slowest setting, DALI remains superior to HybriMoE (e.g., 1.76 
v.s. 1.25 Tokens/s on DeepSeek).
Finally, we select \textbf{(4,\,8)} for Qwen and DeepSeek, and
\textbf{(4,\,1)} 
for Mixtral.
Notably, even with the slowest setting, DALI remains superior to HybriMoE (e.g., 1.76 
v.s. 1.25 Tokens/s on DeepSeek).

\begin{figure}[t]
  \centering
  \vspace{-0.4cm}
  \includegraphics[scale=0.32,trim=0cm 0cm 0cm 0cm,clip]{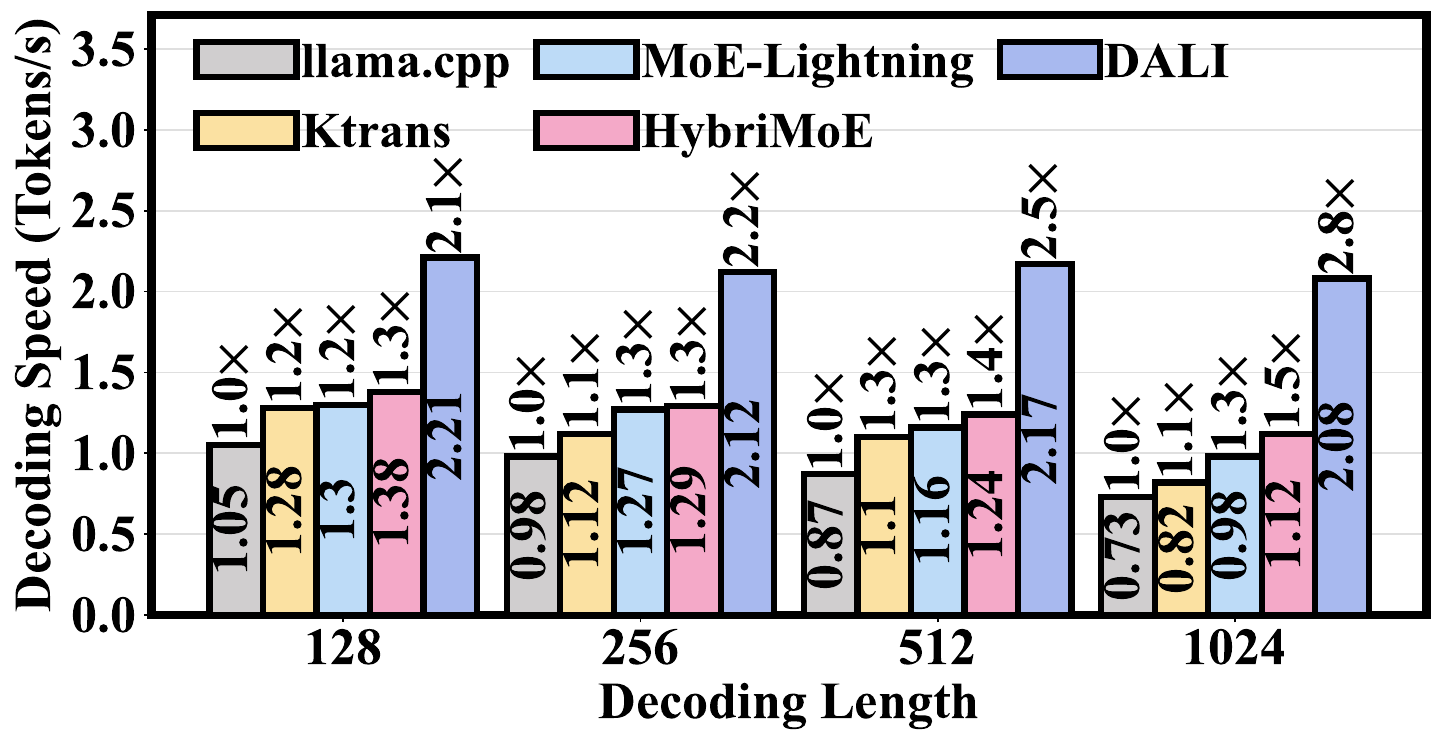}
  \vspace{-0.2cm}
  \caption{Decoding speed on DeepSeek 
  under varying decoding lengths.}
  \vspace{-0.4cm}
  \label{various_decode_length}
\end{figure}

\subsection{Performance under Varying Decoding Lengths}
\label{various_decode_len_appendix}

As shown in Figure~\colorref{various_decode_length}, we report the decoding speed 
under different decoding lengths. 
The experimental results demonstrate that \textsc{DALI} consistently outperforms prior methods across all 
decoding lengths.

\bibliographystyle{ACM-Reference-Format}

\bibliography{references}

\end{document}